\documentclass[12pt]{iopart}
\usepackage{amssymb}
\usepackage{amsfonts}
\usepackage{psfrag}
\usepackage{graphicx}
\usepackage{color}
\usepackage{cite}
\eqnobysec

\def\keywords#1{\vspace{10pt}
     \begin{indented}
     \item[]\rm Keywords: #1\par
     \end{indented}}
\def\be{\begin{equation}}
\def\ee{\end{equation}}
\def\bea{\begin{eqnarray}}
\def\eea{\end{eqnarray}}

\def\CL{{\mathcal L}}
\def\CT{{\mathcal T}}

\def\CD{{\mathcal D}}
\def\fT{{\mathfrak T}}
\def\fU{{\mathfrak U}}
\def\fQ{{\mathfrak Q}}

\def\fD{{\mathfrak P}}

\def\lc{\stackrel{\rm lcsl}{=}}
\def\ic{\stackrel{\rm icsl}{=}}
\def\ev{\stackrel{\rm evsl}{=}}
\def\ov{\overline}
\def\ul{\underline}
\def\al{\alpha}
\def\bet{\beta}

\def\pa{\partial}
\def\ord{\mathrm O}

\begin{document}
\jl{1}

\title[Random sequential adsorption of k-mers]{Random sequential adsorption of k-mers on the fully-connected lattice: probability distributions of the covering time and extreme value statistics}
\author{Lo\"\i c Turban}

\address{Laboratoire de Physique et Chimie Th\'eoriques, Universit\'e de Lorraine--CNRS (UMR7019),
Vand\oe{u}vre l\`es Nancy Cedex, F-54506, France} 

\ead{loic.turban@univ-lorraine.fr}

\begin{abstract}
We study the random sequential adsorption of $k$-mers on the fully-connected 
lattice with $N=kn$ sites. The probability distribution $T_n(s,t)$ of the time $t$ needed to cover the lattice with $s$ $k$-mers is obtained using a generating function approach. In the low coverage scaling limit where $s,n,t\to\infty$ with $y=s/n^{1/2}=\ord(1)$ the random variable $t-s$ follows a Poisson distribution with mean $ky^2/2$. In the intermediate coverage scaling limit, when both $s$ and $n-s$ are $\ord(n)$, the mean value and the variance of the covering time are growing as $n$ and the fluctuations are Gaussian. When  full coverage is approached the scaling functions diverge, which is the signal of a new scaling behaviour. Indeed, when $u=n-s=\ord(1)$, the mean value of the covering time grows as $n^k$ and the variance as $n^{2k}$, thus $t$ is strongly fluctuating and no longer self-averaging. In this scaling regime the 
fluctuations are governed, for each value of $k$, by 
a different extreme value distribution, indexed by $u$. Explicit results are obtained for monomers (generalized Gumbel distribution) and dimers.  
 
\end{abstract}




\section{Introduction} 
{\it Random sequential adsorption} (RSA) is one of the simplest model
of irreversible process in which particles of various shapes 
are deposited randomly on a substrate, one at a time (see \cite{bartelt91,evans93,talbot00,krapivsky10} for reviews). The process takes place either on a lattice or in the continuum. Overlap is forbidden and once deposited a particle remains fixed forever. It follows that, on a finite system, after some time there is not enough place left on the surface to add a new particle: a {\it jammed configuration} is reached. 

The RSA model is expected to describe the adsorption of proteins, colloids or macromolecules 
on homogeneous substrates for which the relaxation time is much longer than the deposition time~\cite{macritchie78,feder80}. The model have also been used in the study of crystal growth, glass formation and reactions on polymer chains~\cite{evans93}.

Quantities of interest are the  jamming fraction, which depends 
on the initial conditions, and the kinetics governing 
the approach to the jammed state. 

Exact results have been mostly obtained in one dimension (1d)~\cite{flory39,renyi58,page59,renyi63,mackenzie62,widom66,mullooly68,widom73,gonzalez74,
hemmer89,monthus91,krapivsky92}. 
For dimer deposition on a lattice, the jamming density is $x_{jam}=1-\e^{-2}=0.864664\ldots$, a result first obtained by Flory in a study of the cyclization reaction along a polymer chain~\cite{flory39}. This limit is approached exponentially in time~\cite{gonzalez74}.

When unit length particles are deposited at arbitrary random positions along a line (continuous RSA), one obtains the {\it car parking problem} solved by R\'{e}nyi~\cite{renyi58,renyi63}. The jamming  
density has a complicated mathematical expression such that 
$x_{jam}=0.747597\ldots$ and the approach to saturation is algebraic, 
namely, $x_{jam}-x(t)\sim t^{-1}$~\cite{renyi58,renyi63,krapivsky10}. 

Analytical results are sparse in higher dimensions and restricted to limited domains of time evolution and density. Numerical studies
suggest that, like in 1d, the approach to jamming is generally exponential in time for deposition on a lattice while it is algebraic on a continuous substrate~\cite{bartelt91,evans93,talbot00,krapivsky10}. In the latter case, the jamming exponent is $1/d$ for isotropic 
objects~\cite{feder80,pomeau80,swendsen81} while it departs from this value for unoriented anisotropic objects~\cite{talbot89,vigil90,viot90,tarjus91,viot92,wang96,baule17}. 

An exception to this lack of exact results
is provided by the deposition of dimers on the Cayley tree with coordination number $z$, a problem which can be solved using the empty connected cluster method~\cite{evans84}.  
The jamming density, $x_{jam}=1-(z-1)^{-z/(z-2)}$~\footnote{Note that this expression gives back Flory's result when $z=2$.}, is approached exponentially in time~\cite{evans84,krapivsky10}.  

In the present work we study the RSA of $k$-mers on the fully connected lattice, a problem for which the time evolution of the density $x(t)$ has been obtained in 1d~\cite{krapivsky10}. 
Here the emphasis is put on the probability distribution of the covering time, its scaling behaviours at low and intermediate coverage and its extreme value statistics in the vicinity of full coverage. This is a continuation of previous works about random walks and reaction-diffusion  processes on the complete graph~\cite{turban14,turban15,turban18a,turban18b}.

Note that for monomers our results remain valid for any homogeneous lattice with $N=n$ sites in any dimension. Furthermore, the RSA of monomers on the fully-connected lattice is equivalent to the covering of the lattice by a random walk~\cite{turban14,turban15} as well as to the 
{\it coupon collector's problem}~\cite{baum65,feller68,flajolet09,boneh97,wiki19}. 

The outline of the paper is as follows. In section 2 the model is described, solved in the mean-field approximation and the main results are presented. Section 3 deals with the probability distribution $T_n(s,t)$ of the time $t$ needed to covers a finite-size lattice with $s$ $k$-mers. An ordinary generating function is written down, leading to a general expression for $T_n(s,t)$ which is afterwards specified for $k$-mers. The same is done for the moments of the distribution. In section 4~\footnote{Section 4 can also be read directly after section 2.}, using a master equation approach, the behaviour of the covering time is studied in three different scaling regimes: first at low coverage 
when $s=\ord(n^{1/2})$, then at intermediate coverage 
when both $s$ and
$n-s=\ord(n)$ and finally in the extreme value limit, when $n-s=\ord(1)$. We conclude in section 5 and give some complements on the calculations in five appendices.

\section{Model, mean field and main results}
\begin{figure}[!t]
\begin{center}
\includegraphics[width=10cm,angle=0]{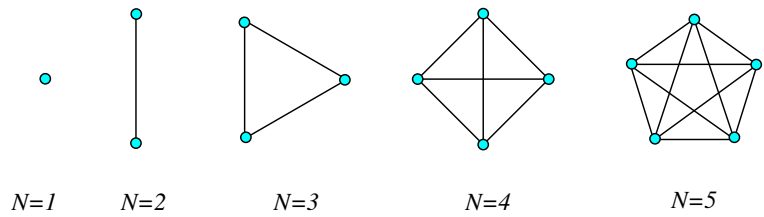}
\end{center}
\vglue -.0cm
\caption{The fully-connected lattice (complete graph) with $N$ sites can be embedded in a Euclidean space of dimension $d=N-1$.
The addition of a new site equidistant from the previous ones requires a new dimension. Therefore all sites belong to the surface.
}\label{fig-1} 
\end{figure}

\subsection{Model}
We study the RSA of $k$-mers on a fully-connected lattice with
$N=kn$ equivalent sites. For such a lattice, by construction, all the sites belong to the surface and can be occupied by $k$-mers (see figure
\ref{fig-1}). Moreover, whatever the value of $k$, the lattice can always be fully covered.  

At each time step $k$ distinct sites are selected at random
among the $N$. Multiple occupancy being forbidden, a new $k$-mer can be adsorbed only when the $k$ selected sites are empty. If $s$ $k$-mers are already covering the lattice, $k(n-s)$ sites among the $kn$ remain free and 
the attempt will be successful with probability
\be
p_n(s)=\frac{{k(n-s)\choose k}}{{kn\choose k}}=\frac{[k(n-s)]^{\ul{k}}}{(kn)^{\ul{k}}}\,,
\label{pns}
\ee
where $j^{\ul{k}}$ is the falling factorial power $j(j-1)\ldots(j-k+1)$~\cite{graham94} and $n$ is the maximum number of adsorbed $k$-mers, corresponding to full coverage.

\subsection{Mean-field theory}

Let $x=s/n$ be the surface coverage, i.e., the fraction 
of lattice sites covered by $k$-mers and $w=t/n$ a rescaled time variable. In the {\it intermediate coverage scaling limit} (icsl) where $s,t,n\to\infty$ with $0<x<1$ and $w=\ord(1)$ the transition probability in~\eref{pns} takes the following form: 
\be
p(x)\ic(1-x)^k\,.
\label{pnx}
\ee
In the mean-field approximation, during the time interval $dt$ 
the number of $k$-mers adsorbed on the lattice changes by 
$ds=ndx=p(x)dt=p(x)ndw$. It follows that:  
\be
\frac{dw}{dx}=\frac{1}{(1-x)^k}\,.
\label{dwdx}
\ee
With the initial condition $w\to0$ when $x\to0$ one obtains:
\be
\frac{t}{n}\ic w=\frac{1}{k-1}\left[\frac{1}{(1-x)^{k-1}}-1\right]\,,
\qquad k>1\,.
\label{wx1}
\ee
The limit $k\to1$ gives the following logarithmic behaviour:
\be
w=-\ln(1-x)\,,\qquad k=1\,.
\label{wx2}
\ee 
Extracting the $k$-mer density from these expressions leads to:
\be
x=\left\{
\begin{array}{ccc}
 1-[1+(k-1)w]^{-1/(k-1)}\,,\qquad &k>1\,,\\
 \ms
1-e^{-w} \,,\qquad &k=1\,.
\end{array}
\right.
\label{xw}
\ee
\subsection{Main results}

The finite-size probability distribution $T_n(s,t)$ 
of the time $t$ needed to cover the lattice with $s$ 
$k$-mers has been obtained as a function of the transition 
probability $p_n(s)$ for any value of $k$ (see~\eref{Tnst3}). 

Depending on the fraction of occupied sites $x=s/n$ three
different scaling regimes are observed when
$n$, $s$ and $t$ tend to infinity:
\begin{itemize}

\item In the {\it low coverage scaling limit} (lcsl), when $y\lc s/n^{1/2}=\ord(1)$ so that $x\lc0$, the probability distribution
$D_n(s,v)=T_n(s,t=s+v)$ converges to the Poisson distribution with mean $\ov{v}=ky^2/2$
\be
\fD_v(y)=\frac{(ky^2/2)^v}{v!}
\exp\left(-\frac{ky^2}{2}\right)\,,\qquad 
\fD_v(0)=\delta_{v,0}\,,
\label{Dvy-0}
\ee
which is the solution of the difference-differential equation
\be
\frac{\pa\fD_v}{\pa y}=ky\left[\fD_{v-1}(y)-\fD_v(y)\right]\,,
\label{pDvy-0}
\ee
following from the master-equation governing $D_n(s,v)$ in this scaling limit.

\item In the {\it intermediate coverage scaling limit} (icsl), when both $s$
and $n-s$ are $\ord(n)$ so that $0<x<1$, the mean covering time
$\ov{t_n}$ and the variance $\ov{\Delta t_n^2}$ are growing as $n$ and $n^{1/2}T_n(s,t)$ converges to the Gaussian probability density
\be
\fT(x,\tau)=\frac{\exp\left[-\frac{\tau^2}{2g(x)}
\right]}{\sqrt{2\pi g(x)}}\,,\qquad \tau\ic\frac{t-\ov{t_n}}{n^{1/2}}\,,
\label{Txtau0}
\ee
where $\ov{t_n}$ is
the mean-field result in~\eref{wx1} and~\eref{wx2}. The variance is given by: 
\be\fl
\frac{\ov{\Delta t_n^2}}{n}\ic g(x)=
\left\{
\begin{array}{ccc}
\frac{1}{2k-1}\left[\frac{1}{(1-x)^{2k-1}}-1\right]-\frac{1}{k-1}\left[\frac{1}{(1-x)^{k-1}}-1\right]
\,,\qquad &k>1\,,\\
 \ms
\frac{x}{1-x}+\ln(1-x)\,,\qquad &k=1\,.
\end{array}
\right.
\label{gtx0}
\ee
Both $\ov{t_n}$ and $\ov{\Delta t_n^2}$ vanish when $x\to0$ and diverge when $x\to1$. 
These divergences indicate a new scaling behaviour at $x=1$

\item In the {\it extreme value scaling limit} (evsl)
when $u=n-s=\ord(1)$ so that $x\ev 1$, $\ov{t_n}$ is growing as 
$(kn)^{\ul{k}}$ and $\ov{\Delta t_n^2}$ as $[(kn)^{\ul{k}}]^2$. Thus the system is non-self-averaging. The appropriate scaling variable is now 
$\tau'=t/(kn)^{\ul{k}}-\delta_{k,1}\ln n$. In the scaling limit, the master equation governing the behaviour of $T_n(s,t)$ leads to the following difference-differential equation 
\be
\frac{\pa\fT'_u}{\pa\tau'}=[k(u+1)]^{\ul{k}}\left[\fT'_{u+1}(\tau')-\fT'_u(\tau')\right]
\label{difTutau0}
\ee
for the extreme value distribution $\fT'_u(\tau')\ev (kn)^{\ul{k}}\,T_n(s,t)$.
The solution for monomers is the generalized Gumbel 
distribution~\cite{baum65,pinheiro16}:
\be
\fT'_u(\tau')=\frac{1}{u!}\exp[-(u+1)\tau'-\e^{-\tau'}]\,.
\label{Tutau0}
\ee
Taking directly the scaling limit on the scaled probability distribution, the following 
expression is obtained for dimers:
\bea
\fl\fT'_u(\tau')=\frac{1}{(2u)!}\sum_{j=0}^\infty(-1)^j
\frac{(4u+4j+3)(4u+2j+1)!!}{(2j)!!}
\,\e^{-(2u+2j+1)(2u+2j+2)\tau'}\,.
\label{Tutau1}
\eea
For each value of $k$ there is a different extreme value distribution of the scaled time, indexed by $u$.

\end{itemize}

\section{Finite-size results}

In this section we study the probability distribution $T_n(s,t)$ of the time $t$ needed to adsorb irreversibly $s$ $k$-mers on the fully-connected lattice with $kn$ sites. 
$T_n(s,t)$ and its moments are obtained using generating functions techniques for general values of $p_n(s)$ before specifying the form of the transition probability for $k$-mers.

\begin{figure}[!t]
\begin{center}
\includegraphics[width=11cm,angle=0]{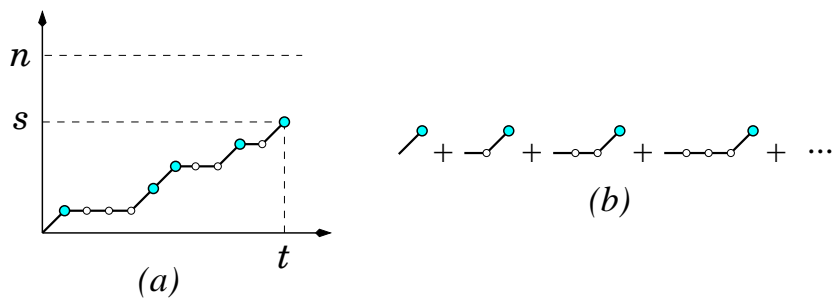}
\end{center}
\vglue -.0cm
\caption{
A time evolution of the number of  $k$-mers 
covering the lattice is sketched in (a). A big circle
corresponds to the successful deposition (probability $p_n(\bet)$) 
of a $k$-mer on the lattice covered by $\bet$ $k$-mers, a small circle to a failed attempt. 
The diagrams in (b) give the contributions to the generating 
function of the probability distribution 
$L_n(\bet,t)$ of the waiting time $t$ in a state with $\bet$ $k$-mers defined in~\protect\eref{Lbetz}.\label{fig-2}  
}
\end{figure}  

\subsection{Generating function for $T_n(s,t)$}
The probability distribution of the covering time $T_n(s,t)$ 
will be deduced from the generating function:
\be
\CT_n(s,z)=\sum_{t=1}^\infty z^t T_n(s,t)\,.
\label{def-Tnsz}
\ee
Let $\bet$ be the intermediate number of $k$-mers deposited on the lattice, its evolution from 0 to $s$ 
in figure~\ref{fig-2}(a)
takes place through sequences where the system remains for some time 
in a state with a constant value of $\beta$, ending with a 
transition, $\bet\to\bet+1$. To the probability distribution of the waiting time, $L_n(\bet,l)$ one can associate the generating function $\CL_n(\bet,z)$ corresponding to the diagrams of figure~\ref{fig-2}(b):   
\bea
\CL_n(\bet,z)&=\!\sum_{l=1}^\infty z^lL_n(\bet,l)\!=\!\{1\!+\!z[1\!-p_n(\bet)]\!+\!z^2[1\!-p_n(\bet)]^2\!+\cdots\}zp_n(\bet)\nonumber\\
&=\frac{zp_n(\bet)}{1-z[1-p_n(\bet)]}\,.
\label{Lbetz}
\eea
The generating function we are looking for is simply obtained 
as the product 
\be
\CT_n(s,z)=\prod_{\bet=0}^{s-1}\CL_n(\bet,z)
=z^s\prod_{\bet=0}^{s-1}\frac{p_n(\bet)}{1-z[1-p_n(\bet)]}\,.
\label{Tnsz}
\ee
Note that $\CT_n(s,1)=\sum_{t=1}^\infty T_n(s,t)=1$ as required.

\subsection{General expression of $T_n(s,t)$}

\begin{figure}[!t]
\begin{center}
\includegraphics[width=9cm,angle=0]{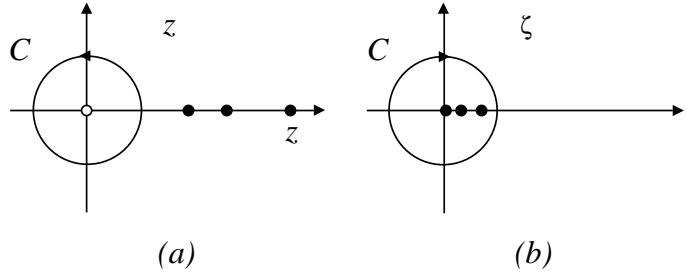}
\end{center}
\vglue -.5cm
\caption{Integration contour $C$ used to extract $T_n(s,t)$ from its generating function in~\eref{Tnst1}. (a) In the z-plane $C$ is the unit circle, centered at the origin and oriented counter-clockwise. The integrand has a pole at the origin when $t\geq s$ and $s$ simple poles outside $C$ on the positive real axis. (b) Through the transformation $\zeta=1/z$, the unit circle transforms into itself, but $C$ is now oriented clockwise. The pole at the origin is sent to infinity 
and the $s$ simple poles are transported inside $C$.
\label{fig-3}
}
\end{figure}

According to \eref{def-Tnsz} the probability distribution, 
which is the coefficient of $z^t$ in the series expansion 
of $\CT_n(s,z)$ in~\eref{Tnsz}, is given by
the following contour integral:
\bea
T_n(s,t)&=\frac{1}{2i\pi}\oint_{\stackrel{\curvearrowleft}{C}} dz\frac{\CT_n(s,z)}{z^{t+1}}
=\frac{1}{2i\pi}\oint_{\stackrel{\curvearrowleft}{C}}\frac{dz}{z^{t-s+1}}
\prod_{\bet=0}^{s-1}\frac{p_n(\bet)}{1-z[1-p_n(\bet)]}\nonumber\\
&=\frac{1}{2i\pi}\oint_{\stackrel{\curvearrowleft}{C}}\frac{dz}{z^{t+1}}
\prod_{\bet=0}^{s-1}\frac{p_n(\bet)}{1/z-[1-p_n(\bet)]}\,.
\label{Tnst1}
\eea
The contour $\stackrel{\curvearrowleft}{C}$ is a unit circle, 
oriented counter-clockwise and centered at the origin. 
Besides the pole at the origin for $t\geq s$, 
the integrand has $s$ simple poles $z_\al=[1-p_n(\al)]^{-1}$ 
on the positive real axis outsides 
$\stackrel{\curvearrowleft}{C}$ since $0<p_n(\al)\leq1$ by assumption. 

As shown in figure \ref{fig-3}, the change 
of variable $\zeta=1/z$ transforms the contour $\stackrel{\curvearrowleft}{C}$
into $\stackrel{\curvearrowright}{C}$, a unit circle with a clockwise orientation.
It sends to infinity the pole at the origin and introduces $s$ simple poles 
$\zeta_\al=[1-p_n(\al)]<1$ inside $\stackrel{\curvearrowright}{C}$ so that:
\be
\fl T_n(s,t)=-\frac{1}{2i\pi}\oint_{\stackrel{\curvearrowright}{C}}d\zeta\,\zeta^{t-1}
\prod_{\bet=0}^{s-1}
\frac{p_n(\bet)}{\zeta\!-\![1\!-\!p_n(\bet)]}
=\frac{1}{2i\pi}\oint_{\stackrel{\curvearrowleft}{C}}d\zeta\,\zeta^{t-1}
\prod_{\bet=0}^{s-1}
\frac{p_n(\bet)}{\zeta\!-\![1\!-\!p_n(\bet)]}\,.
\label{Tnst2}
\ee

Finally, applying the theorem of residues, one obtains
the probability distribution
\be
T_n(s,t)=\prod_{\bet=0}^{s-1}p_n(\bet)\sum_{\al=0}^{s-1}
\frac{[1-p_n(\al)]^{t-1}}{\prod_{\bet=0\atop \bet\neq\al}^{s-1}[p_n(\bet)-p_n(\al)]}\,.
\label{Tnst3}
\ee

\subsection{Explicit expression of $T_n(s,t)$ for $k$-mers}
It will be convenient to express $T_n(s,t)$ as a function 
of $u=n-s$ in order to
study the scaling limit when $u=\ord(1)$. According to~\eref{pns}
the probability distribution in~\eref{Tnst3} can be rewritten as:
\bea
\fl&T_n(s,t)=\sum_{\al=0}^{s-1}\frac{ab}{cd}\,p_n(\al)
[1-p_n(\al)]^{t-1}\,,
\quad a=\prod_{\bet=0}^{\al-1}[k(n-\bet)]^{\ul{k}}\,,\quad 
b=\prod_{\bet=\al+1}^{s-1}[k(n-\bet)]^{\ul{k}}\,,\nonumber\\
\fl&c=\prod_{\bet=0}^{\al-1}\left\{[k(n-\bet)]^{\ul{k}}-
[k(n-\al)]^{\ul{k}}\right\}\,,\quad 
d=\prod_{\bet=\al+1}^{s-1}\left\{[k(n-\bet)]^{\ul{k}}-
[k(n-\al)]^{\ul{k}}\right\}\,.
\label{Tabcd}
\eea
With the change of variables 
\be
\al=s-j-1=n-u-j-1\,,\qquad\bet=s-i-1=n-u-i-1\,,
\label{albet}
\ee
one obtains
\bea
\fl&c\!=\!\!\!\prod_{i=j+1}^{n-u-1}\!\!\left\{[k(u\!+\!i\!+\!1)]^{\ul{k}}\!-\!
[k(u\!+\!j\!+\!1)]^{\ul{k}}\right\}\,,\quad
d\!=\!(-1)^j\prod_{i=0}^{j-1}\left\{[k(u\!+\!j\!+\!1)]^{\ul{k}}\!-\!
[k(u\!+\!i\!+\!1)]^{\ul{k}}\right\}\,,\nonumber\\
\fl&a=\frac{(kn)!}{[k(u\!+\!j+\!1)]!}\,,\quad 
b=\frac{[k(u\!+\!j)]!}{(ku)!}\,,\quad 
ab\,p_n(\al)=\frac{(kn)!}{(kn)^{\ul{k}}(ku)!}\,,
\label{abcd}
\eea
so that:
\be
T_n(n-u,t)=\frac{(kn)!}{(kn)^{\ul{k}}(ku)!}\sum_{j=0}^{n-u-1}
\frac{1}{cd}\left\{1-\frac{[k(u\!+\!j+\!1)]^{\ul{k}}}{(kn)^{\ul{k}}} \right\}^{t-1}\,.
\label{Tnut}
\ee
When $k=1$, the probability distribution has a
simple expression~\cite{turban15} in terms of Stirling's numbers of the second kind~\cite{stirling49}: 
\be
T_n(s,t)=\frac{n^{\ul{s}}}{n^t}{t-1\brace s-1}\,,\qquad
{t\brace s}=\frac{1}{s!}
\sum_{j=0}^{s}(-1)^{s-j}{s\choose j}j^{t}\,.
\label{Tnstir-1}
\ee

\subsection{Moments of the probability distribution}
Since the covering time $t(s)$ is the sum from $\bet=0$ to $s-1$ of the waiting times, $l(\bet)$,
which are independent random  variables, its mean value and variance are given by the corresponding sums for the waiting times.

\subsubsection{Mean value.}
The mean value of the waiting time in a state with $\bet$ adsorbed $k$-mers is given by the first derivative at $z=1$ of the generating function $\CL_n(\bet,z)$ 
in~\eref{Lbetz}
\be
\ov{l_n(\bet)}=\sum_{l=1}^\infty l\,L_n(\bet,l)=\left.\frac{\pa\CL_n}{\pa z}\right|_{z=1}
=\frac{1}{p_n(\bet)}\,,
\label{lnb}
\ee
so that:
\be
\ov{t_n(s)}=\sum_{\bet=0}^{s-1}\frac{1}{p_n(\bet)}\,.
\label{tns}
\ee

\subsubsection{Variance.}
A second derivative gives the mean square value
\be
\ov{l_n^2(\bet)}=\sum_{l=1}^\infty l^2L_n(\bet,l)=\left.\frac{\pa}{\pa z}\left[z
\frac{\pa\CL_n}{\pa z}\right]\right|_{z=1}
=\frac{2-p_n(\bet)}{p_n^2(\bet)}\,,
\label{lnb2}
\ee
from which follows the variance:
\be
\ov{\Delta l_n^2(\bet)}=\ov{l_n^2(\bet)}-\ov{l_n(\bet)}^2
=\frac{1-p_n(\bet)}{p_n^2(\bet)}=\frac{1}{p_n^2(\bet)}-\ov{l_n(\bet)}\,.
\label{dlnb2}
\ee
Thus the variance of the covering time is given by:
\be
\ov{\Delta t_n^2(s)}=\sum_{\bet=0}^{s-1}\frac{1}{p_n^2(\bet)}-\ov{t_n(s)}\,.
\label{dtns2}
\ee

\subsection{Explicit expressions of the moments for k-mers}
\subsubsection{Mean value.}
The mean value of the covering time, following from~\eref{tns}
and~\eref{pns}, is given by:
\be
\ov{t_n(s)}=\sum_{\beta=0}^{s-1}\frac{(kn)^{\ul{k}}}{[k(n-\beta)]^{\ul{k}}}=\sum_{j=n-s+1}^n\frac{(kn)^{\ul{k}}}{(kj)^{\ul{k}}}\,.
\label{tnsk-1}
\ee
Using~\eref{aka} in appendix~A, one obtains:
\be
\ov{t_n(s)}=\frac{(kn)^{\ul{k}}}{(k-1)!}\sum_{j=n-s+1}^n
\sum_{l=0}^{k-1}(-1)^{k-l-1}\frac{{k-1\choose l}}{kj-l}\,.
\label{tnsk-2}
\ee
As shown in appendix~B, this expression can be rewritten 
using harmonic and generalized harmonic numbers, under a
form which is appropriate 
to obtain the scaling limit when $u=n-s=\ord(1)$:   
\bea\fl
\ \ \ov{t_n}&=\frac{(kn)^{\ul{k}}}{(k-1)!}
\left[
H_{kn}-H_{ku}+(\delta_{k,1}-1)(H_n-H_u)
+\sum_{m=2}^\infty\frac{H_n^{(m)}-H_u^{(m)}}{k^{m}}\;P_{k,m}
\right]\,,\nonumber\\
\fl P_{k,m}&=\sum_{l=0}^{k-1}\left[(-1)^{k-l-1}{k-1\choose l}-1\right]l^{m-1}\,.
\label{tnsk-3} 
\eea
For monomers and dimers this expression reduces to:
\bea
k&=1\,,\quad\ov{t_n}&=n\,(H_n-H_u)\,,\nonumber\\
k&=2\,,\quad\ov{t_n}&=(2n)^{\ul{2}}\,(H_{2n}-H_{2u}-H_n+H_u)\,.\label{tn12}
\eea

\subsubsection{Variance.}
Using~\eref{pns} and~\eref{aka} the first contribution to the variance in~\eref{dtns2} is given by:
\be\fl
\sum_{\bet=0}^{s-1}\frac{1}{p_n^2(\bet)}
=\ov{\Delta t_n^2(s)}+\ov{t_n(s)}=\left[\frac{(kn)^{\ul{k}}}{(k-1)!}\right]^2
\sum_{j=n-s+1}^n\sum_{l,l'=0}^{k-1}(-1)^{l+l'}
\frac{{k-1\choose l}{k-1\choose l'}}{(kj-l)(kj-l')}\,.
\label{var-1}
\ee
As above for the mean value, this expression can be rewritten in
terms of generalized harmonic numbers (see 
appendix~C):
\bea
\fl\ov{\Delta t_n^2}\!+\ov{t_n}&=\left[\frac{(kn)^{\ul{k}}}{(k-1)!}\right]^2
\left\{
H_{kn}^{(2)}\!-\!H_{ku}^{(2)}\!+\!\sum_{m=2}^\infty
\frac{H_n^{(m)}\!-\!H_u^{(m)}}{k^m}
\left[
(m\!-\!1)Q_{k,m}\!+\!2R_{k,m}
\right]
\right\}\,,\nonumber\\
\fl\quad\  Q_{k,m}&=\sum_{l=0}^{k-1}\left[{k-1\choose l}^2-1\right]l^{m-2}\,,
\nonumber\\
\fl\quad\  R_{k,m}&=\sum_{l>l'=0}^{k-1}(-1)^{l+l'}{k-1\choose l}
{k-1\choose l'}\frac{l^{m-1}-{l'}^{m-1}}{l-l'}\,.
\label{var-2}
\eea
Thus for monomers and dimers, taking~\eref{tn12} into account, 
the variance is given by:
\bea
\fl &k=1\,,\quad\ov{\Delta t_n^2}=n^2[H_n^{(2)}-H_u^{(2)}]
-n(H_n-H_u)\,,\nonumber\\
\fl &k=2\,,\quad\ov{\Delta t_n^2}=[(2n)^{\ul{2}}]^2
[H_{2n}^{(2)}-H_{2u}^{(2)}]-(2n)^{\ul{2}}[2(2n)^{\ul{2}}+1]
(H_{2n}-H_{2u}-H_n+H_u)\,.
\label{dtn212}
\eea

\section{Scaling limits}
In this section we study the different scaling limits of the problem using a master-equation approach~\footnote{For the coupon collector's problem, corresponding to $k=1$, these limits have been taken directly on the generating 
function~\cite{baum65}.}.

\subsection{Master equations}
Let $S_n(s,t)$ be the probability to have $s$ k-mers covering $ks$ sites on the lattice at time $t$ and 
$p_n(s)$ the probability to add a k-mer on the lattice with $s$ k-mers. Between $t$ and $t+1$, $s\rightarrow s$ with probability $1-p_n(s)$ and
$s-1\rightarrow s$ with probability $p_n(s-1)$ thus $S_n(s,t)$ evolves according to the following master equation:
\be
S_n(s,t+1)=[1-p_n(s)]S_n(s,t)+p_n(s-1)S_n(s-1,t).
\label{master-Sn}
\ee
The probability distribution of the time $t$ needed to adsorb $s$ $k$-mers (first-passage time by the value $s$) is given by:
\be 
T_n(s,t)=S_n(s-1,t-1)p_n(s-1)\,. 
\label{def-Tn}
\ee
According to~\eref{master-Sn}, its time evolution is governed by the following master equation:
\be
T_n(s,t+1)=[1-p_n(s-1)]T_n(s,t)+p_n(s-1)T_n(s-1,t).
\label{master-Tn}
\ee
Changing $t$ into $t-1$ in~\eref{master-Tn} the 
probability distribution $D_n(s,v)=T_n(s,t)$ 
with $v=t-s$ evolves as follows:
\be
D_n(s,v)=[1-p_n(s-1)]D_n(s,v-1)+p_n(s-1)D_n(s-1,v).
\label{master-Dn}
\ee

\subsection{Low coverage scaling limit}

\begin{figure}[!t]
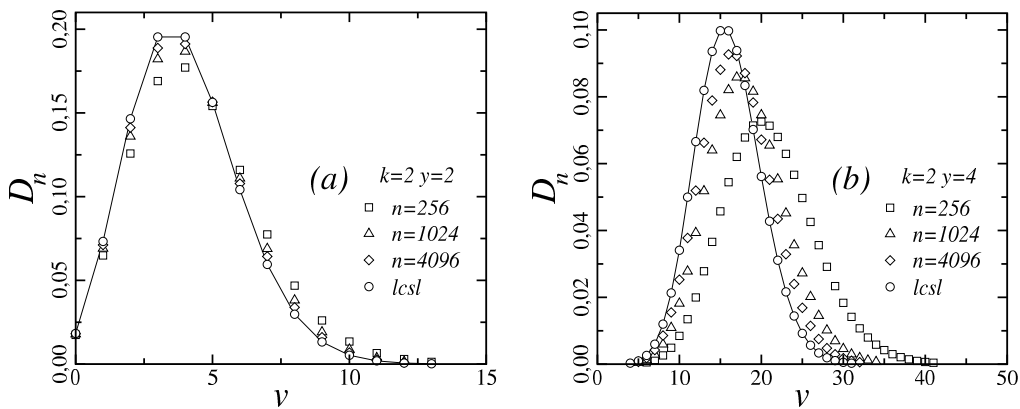

\begin{center}
\includegraphics[width=6.5cm,angle=0]{fig-4a.eps}
\hglue 3mm
\includegraphics[width=6.5cm,angle=0]{fig-4b.eps}
\end{center}
\vglue -.5cm
\caption{(a) Finite-size data collapse on the Poisson distribution $\fD_v(y)\lc D_n(s,v)$ in~\eref{Dvy-4} versus $v=t-s$ when $k=2$ and $y=s/n^{1/2}=2$. (b) When $y=4$, 
for the same sizes, the convergence to the limit 
distribution is much slower. The system crosses over from
Poisson to Gauss when $n$ decreases ($x=y/n^{1/2}$). 
\label{fig-4}
}
\end{figure}

In the lcsl, first considered 
in~\cite{baum65} for the coupon collector's problem, 
$s,t,n\to\infty$  
with $s=\ord(n^{1/2})$ so that:  
\be
y\lc \frac{s}{n^{1/2}}=\ord(1)\,,\qquad 
x=\frac{s}{n}=\frac{y}{n^{1/2}}\lc 0\,. 
\label{yv}
\ee
Let us look for the form of the master 
equation~\eref{master-Dn}
in this scaling limit where we define:
\be
\fD_v(y)=\fD_v[y(s)]\lc D_n(s,v)\,,\qquad v\in\mathbb{N}_0\,.
\label{Dvy-1}
\ee
Since $y(s-1)=y-n^{-1/2}$, a first order expansion gives:
\be
\fD_v[y(s-1)]=\fD_v(y)-\frac{1}{n^{1/2}}
\frac{\pa\fD_v}{\pa y}+\ord(n^{-1})\,.
\label{expDvy}
\ee
According to~\eref{pns}, one has:
\be\fl
p_n(s-1)=\prod_{j=0}^{k-1}
\left[\frac{k(n-s+1)-j}{kn-j}\right]
=\left(1-\frac{y}{n^{1/2}}\right)^k+\ord(n^{-1})
=1-\frac{ky}{n^{1/2}}+\ord(n^{-1})\,.
\label{pnsm}
\ee
Thus the master equation~\eref{master-Dn} can be rewritten as:
\be
\fD_v(y)=\frac{ky}{n^{1/2}}\fD_{v-1}(y)
+\left(1-\frac{ky}{n^{1/2}}\right)
\left[\fD_v(y)-\frac{1}{n^{1/2}}
\frac{\pa\fD_v}{\pa y}\right]+\ord(n^{-1})\,.
\label{Dvy-2}
\ee
The leading contribution, of order $n^{-1/2}$, gives 
the following difference-differential equation for $\fD_v(y)$:
\be
\frac{\pa\fD_v}{\pa y}=ky\left[\fD_{v-1}(y)-\fD_v(y)\right]\,.
\label{pDvy-1}
\ee
Making use of the generating function
\be
\CD(\lambda,y)=\sum_{v=0}^\infty \lambda^v\fD_v(y)\,,
\label{Dly-1}
\ee
equation~\eref{pDvy-1} transforms into:
\be
\frac{\pa\CD}{\pa y}=ky(\lambda-1)\CD(\lambda,y)\,.
\label{pDly}
\ee
The solution satisfying $\CD(1,y)=1$ reads
\be
\CD(\lambda,y)=\exp\left[(\lambda-1)\frac{ky^2}{2}\right]\,,
\label{Dly-2}
\ee
which is the generating function of the Poisson distribution
given by
\be
\fD_v(y)=\frac{(ky^2/2)^v}{v!}
\exp\left(-\frac{ky^2}{2}\right)\,,
\label{Dvy-4}
\ee
in agreement with~\cite{baum65} when $k=1$.
The convergence of the finite-size data, obtained by iterating~\ref{master-Dn}, to the Poisson distribution is shown in figure~\ref{fig-4}.
Since $s$ is non-fluctuating, one obtains: 
\be
\ov{v}= ky^2/2\lc\ov{t_n}-s\,,\qquad\ov{\Delta v^2}
=ky^2/2\lc\ov{\Delta t_n^2}\,. 
\label{mv}
\ee
These results can be verified using~\eref{pns} 
in~\eref{tns} and~\eref{dtns2}, in the same limit.
When $y=0$, i.e., when $s=\ord(n^\alpha)$ with 
$\alpha<1/2$, $\fD_v(0)=\delta_{v.0}$.

\subsection{Intermediate coverage scaling limit}
\label{scalTn}

\begin{figure}[!t]
\begin{center}
\includegraphics[width=8cm,angle=0]{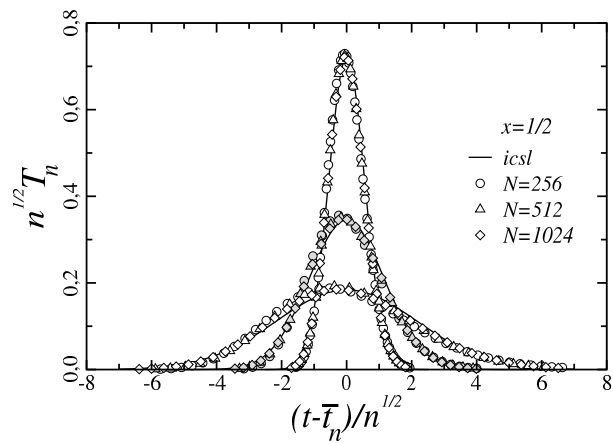}
\end{center}
\vglue -.5cm
\caption{Finite-size data collapse on the Gaussian density 
$\fT\ic n^{1/2}T_n$ in~\eref{Txtau-2} versus $\tau=(t-\ov{t_n})/n^{1/2}$ for $k=1,2,3$
from top to bottom. The fraction of occupied sites is $x=s/n=1/2$. The finite-size results were obtained by 
iterating~\eref{master-Tn}.
\label{fig-5} 
}
\end{figure}

In this section we look for the behaviour of $T_n(s,t)$ in the intermediate coverage scaling limit (icsl), where both $s$ and $u=n-s$ are $\ord(n)$, so that $0<x<1$ when $n\to\infty$. 

The mean-field results~\eref{wx1}
and~\eref{wx2} suggest that $\ov{t_n}$ is scaling as $n$ and a similar behaviour is expected for the variance so that we write:
\be
\frac{\ov{t_n}}{n}\ic f(x)\,,\qquad 
\frac{\ov{\Delta t_n^2}}{n}\ic g(x)\,.
\label{ftgt}
\ee
The appropriate time variable may then be defined as 
(see figure~\ref{fig-5})
\be
\tau(s,t)\ic\frac{t-\ov{t_n}}{n^{1/2}}
=\frac{t}{n^{1/2}}-n^{1/2}f\left(\frac{s}{n}\right)\,,
\label{tau}
\ee
with the associated probability density given by:
\be
\fT(x,\tau)=\fT[x(s),\tau(s,t)]\ic n^{1/2}T_n(s,t)\,.
\label{Txtau-0}
\ee

\begin{figure}[!t]
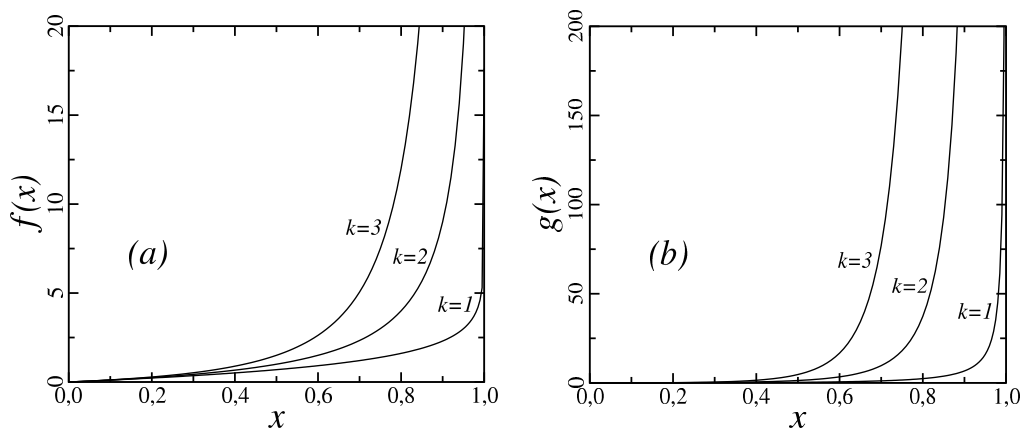

\begin{center}
\includegraphics[width=6.5cm,angle=0]{fig-6a.eps}
\hglue 3mm
\includegraphics[width=6.5cm,angle=0]{fig-6b.eps}
\end{center}
\vglue -.5cm
\caption{Scaling functions (a) $f(x)\ic\ov{t_n}/n$ and 
(b) $g(x)\ic\ov{\Delta t_n^2}/n$ versus $x=s/n$ for $k=1,2,3$. The two functions 
diverge when $x\to1$, which is the signal of a different scaling behaviour in this limit. Note the change of the vertical scale.
\label{fig-6}
}
\end{figure}

Proceeding as above (see appendix~D), at order $n^{-1/2}$ the expansion of the master equation~\eref{master-Tn} gives  
a differential equation~\eref{dftdx} for $f(x)$ such that, with the initial condition $f(0)=0$
\be
f(x)\ic\frac{\ov{t_n}}{n}=
\left\{
\begin{array}{ccc}
\frac{1}{k-1}\left[\frac{1}{(1-x)^{k-1}}-1\right] \,,\qquad &k>1\\
 \ms
-\ln(1-x)\,,\qquad &k=1
\end{array}
\right.\,,
\label{ftx}
\ee
in agreement with the mean-field results~\eref{wx1} 
and~\eref{wx2}. This scaling function is shown in figure~\ref{fig-6}(a). When $x\to0$, $f(x)\sim x$, i.e., each deposition attempt is successful, as expected. When $x\to1$,
$f(x)$ diverges logarithmically for $k=1$ and as 
$(1-x)^{-(k-1)}$ for $k>1$.

The terms of order $n^{-1}$ provide the partial differential equation~\eref{pdeT} for $\fT(x,\tau)$.
Assuming that $\fT(x,\tau)$ depends on $x$ through $g$, the scaling function of the variance, i.e. 
\be
\fT(x,\tau)=\fQ[g(x),\tau]\,,
\label{Txtau-1}
\ee
the partial differential equation can be rewritten as:
\be
\frac{\pa\fQ}{\pa g}\frac{dg}{dx}=\frac{1}{2}\underbrace{\left[
\frac{1}{(1-x)^{2k}}-\frac{1}{(1-x)^k}\right]}_{dg/dx}
\frac{\pa^2\fQ}{\pa\tau^2}\,.
\label{pdeQ}
\ee
The standard diffusion equation is obtained when 
\be
\frac{dg}{dx}=\left[\frac{1}{(1-x)^{2k}}-\frac{1}{(1-x)^k}\right]\,,
\label{dgtdx}
\ee
so that:
\be
\fT(x,\tau)=\fQ[g(x),\tau]=\frac{\exp\left[-\frac{\tau^2}{2g(x)}
\right]}{\sqrt{2\pi g(x)}}\,.
\label{Txtau-2}
\ee
The solution of \eref{dgtdx} is immediate and given by
\be\fl
g(x)\ic\frac{\ov{\Delta t_n^2}}{n}=
\left\{
\begin{array}{ccc}
\frac{1}{2k-1}\left[\frac{1}{(1-x)^{2k-1}}-1\right]-\frac{1}{k-1}\left[\frac{1}{(1-x)^{k-1}}-1\right]
\,,\qquad &k>1\\
 \ms
\frac{x}{1-x}+\ln(1-x)\,,\qquad &k=1
\end{array}
\right.\,.
\label{gtx}
\ee
for $g(0)=0$. The behaviour of this scaling function 
is shown in figure~\ref{fig-6}(b). 
For any value of $k$, $g(x)$ vanishes as 
$kx^2/2$ when $x\to0$ and diverges as $(1-x)^{-(2k-1)}$ 
when $x\to1$. 

The cross-over from Poisson to Gauss at low density is discussed in appendix~E. The divergence of $f(x)$ and $g(x)$ when $x\to1$
indicates a different scaling behaviour when $u=n-s=\ord(1)$. This extreme value limit is studied in the next section.

\subsection{Extreme value scaling limit}

Let us finally study the  approach to saturation when $s, t, n\to\infty$ with $u=n-s=\ord(1)$ so that $x=1-u/n\ev 1$. 
In this extreme-value regime, strong fluctuations of the covering time and anomalous scaling behaviour are expected. 

\subsubsection{Mean value.}
In the limit where $n\to\infty$,
according to~\eref{tnsk-3}, one obtains
\be\fl
\frac{\ov{t_n}}{(kn)^{\ul{k}}}-\delta_{k,1}\ln n\!\ev\!\frac{1}{(k\!-\!1)!}
\left[
\ln k\!+\!\delta_{k,1}\gamma\!-\!H_{ku}\!
+\!(1\!-\!\delta_{k,1})\!H_u\!
+\!\sum_{m\!=\!2}^\infty\!\frac{\zeta(m)\!
-\!H_u^{(m)}}{k^{m}}P_{k,m}
\right]\!,
\label{scaltnsk}
\ee
where $\gamma=0.577\, 215\, 665\,\ldots$ is the Euler-Mascheroni constant, $\zeta(m)$ is the Riemann zeta function and $P_{k,m}$
is defined in~\eref{tnsk-3}.
For monomers and dimers~\eref{tn12} leads to:
\bea
k&=1\,,\quad\frac{\ov{t_n}}{n}-\ln n\ev\gamma-H_u\,,\nonumber\\
k&=2\,,\quad\frac{\ov{t_n}}{(2n)^{\ul{2}}}\ev
\ln 2-H_{2u}+H_u\,.
\label{scaltn12}
\eea

\subsubsection{Variance.}
For the variance~\eref{var-2} gives
\bea
\fl\frac{\ov{\Delta t_n^2}}{\left[(kn)^{\ul{k}}\right]^2}&\!\ev\frac{1}{[(k-1)!]^2}
\left\{
\zeta(2)\!-\!H_{ku}^{(2)}\!+\!\sum_{m=2}^\infty
\frac{\zeta(m)\!-\!H_u^{(m)}}{k^m}
\left[(m\!-\!1)Q_{k,m}\!+\!2R_{k,m}\right]
\right\}\,,
\label{scalvar}
\eea
with $Q_{k,m}$ and $R_{k,m}$ as  defined in~\eref{var-2}. For monomers and dimers~\eref{dtn212}
leads to:
\bea
k&=1\,,\quad\frac{\ov{\Delta t_n^2}}{n^2}\ev\zeta(2)-H_u^{(2)}\,,\nonumber\\
k&=2\,,\quad\frac{\ov{\Delta t_n^2}}{[(2n)^{\ul{2}}]^2}
\ev\zeta(2)-H_{2u}^{(2)}-2(\ln 2-H_{2u}+H_u)\,.
\label{vardtn212}
\eea

\subsubsection{Master equation in the extreme value scaling limit.}

\begin{figure}[!t]
\begin{center}
\includegraphics[width=6.5cm,angle=0]{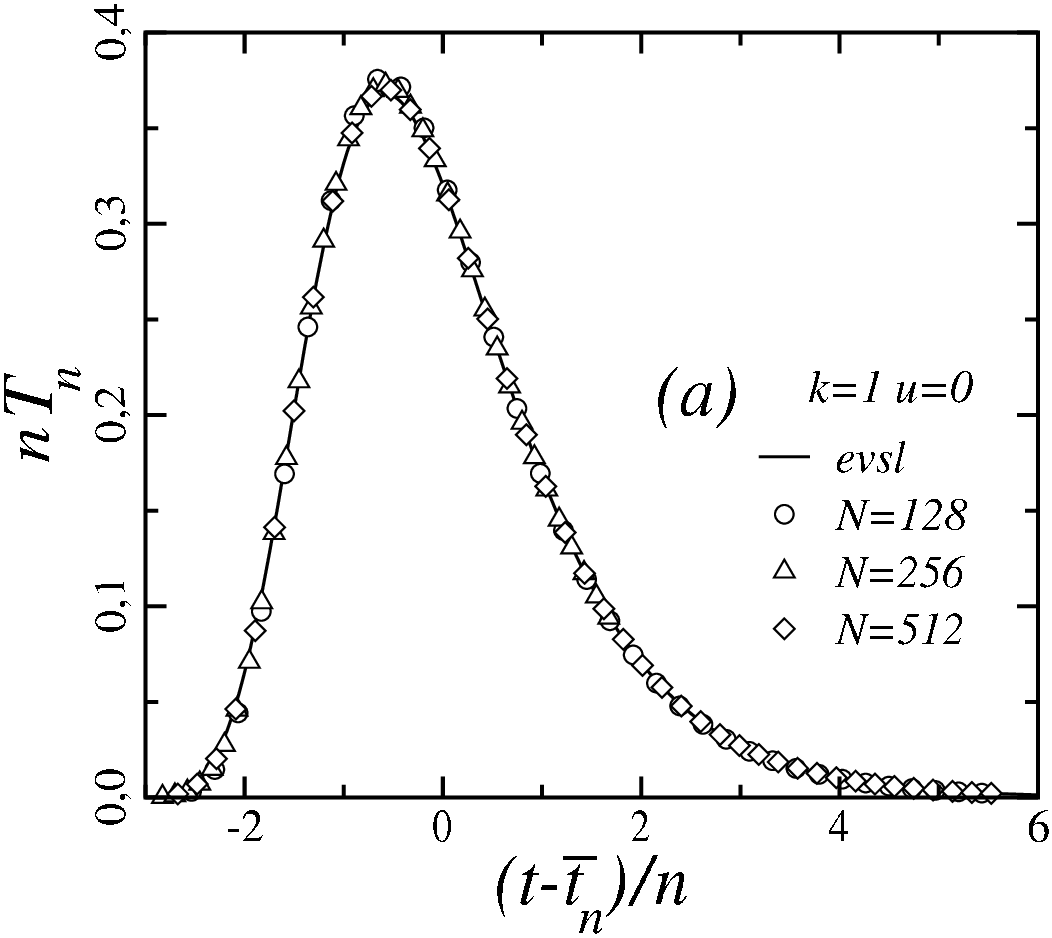}
\hglue 3mm
\includegraphics[width=6.5cm,angle=0]{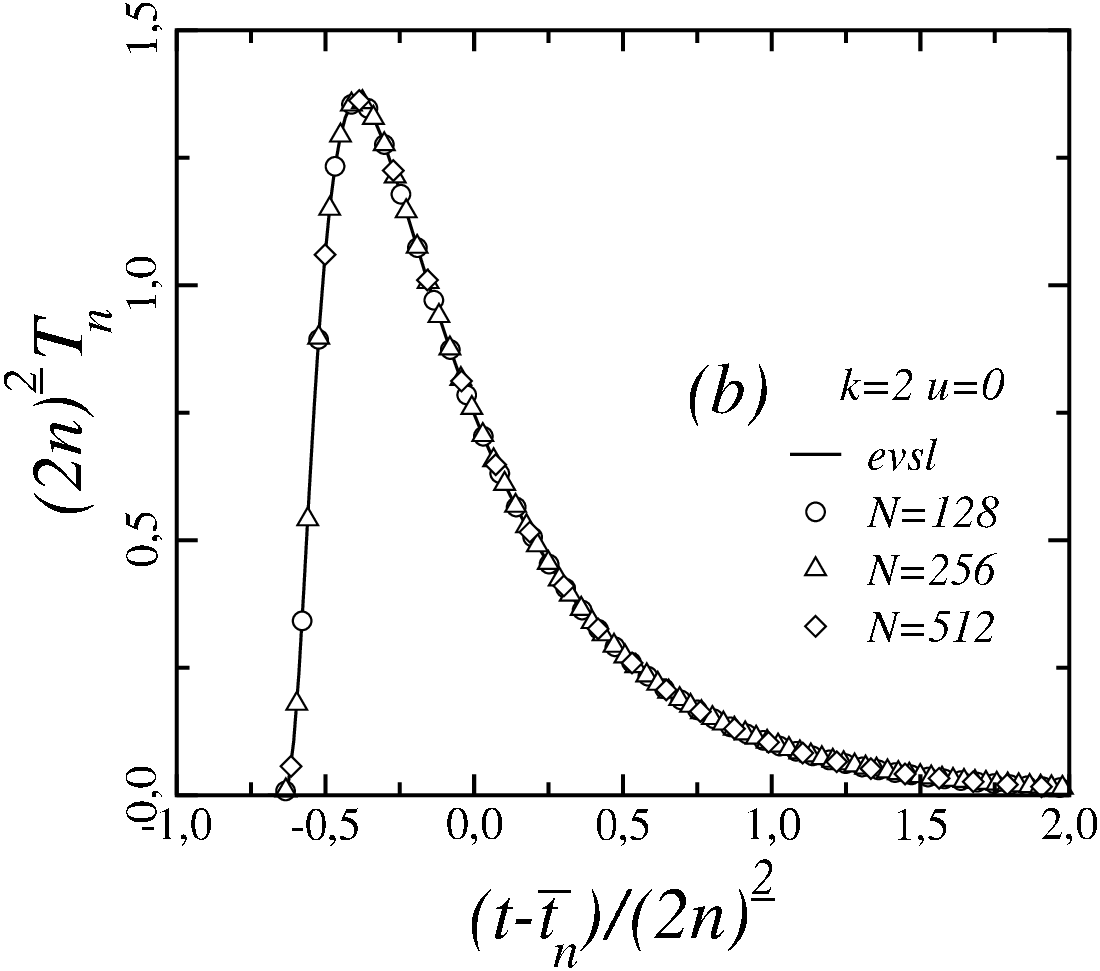}
\end{center}
\vglue -.5cm
\caption{Finite-size data collapse on (a) the Gumble distribution $\fT'_0(\tau')\ev nT_n(n,t)$ in~\eref{Tutau-3} versus $\tau'-\ov{\tau'}=(t-\ov{t_n})/n$ when $k=1$ and $u=0$, (b) the probability density
$\fT'_0(\tau')\ev (2n)^{\ul{2}}T_n(n,t)$ in~\eref{Tutau-4}
versus $\tau'-\ov{\tau'}=(t-\ov{t_n})/(2n)^{\ul{2}}$ when $k=2$ and $u=0$. The finite-size data were obtained by iterating~\eref{master-Tn}.
\label{fig-7}
}
\end{figure}
According to~\eref{scaltnsk} and~\eref{scalvar}, $\ov{t_n}$ and $\sqrt{\ov{\Delta t_n^2}}$
are both growing as~$(kn)^{\ul{k}}$ when $u=\ord(1)$.
Thus, in the scaling limit, we define a new time variable \footnote{Note that for $k=1$, according to~\eref{scaltn12},  
$\tau'$ has to be shifted by $-\ln n$ in order to keep a 
finite mean value in the scaling limit. This shift 
does not affect the results of the present section.}
\be
\tau'=\frac{t}{(kn)^{\ul{k}}}-\delta_{k,1}\ln n
\label{tauk}
\ee
to which corresponds 
the probability density (see figure~\ref{fig-7})
\be
\fT'_u(\tau')\ev (kn)^{\ul{k}}T_n(s,t)\,,\qquad u=n-s=\ord(1)\,.
\label{Tutau-1}
\ee
In the master equation~\eref{master-Tn}, the transition probability is given by:
\be
p_n(s-1)=\frac{[k(n-s+1)]^{\ul{k}}}{(kn)^{\ul{k}}}
\ev\frac{[k(u+1)]^{\ul{k}}}{(kn)^k}\,.
\label{pnu}
\ee
A first-order Taylor expansion of $\fT'_u\left[\tau'+(kn)^{-k}\right]$ on the left-hand-side of the master equation~\eref{master-Tn} leads to:
\be\fl
\fT'_u(\tau')\!+\!\frac{1}{(kn)^k}\frac{\pa\fT'_u(\tau')}{\pa\tau'}\!
=\!\left\{\!1\!-\frac{[k(u\!+\!1)]^{\ul{k}}}{(kn)^k}\right\}\fT'_u(\tau')
\!+\!\frac{[k(u\!+\!1)]^{\ul{k}}}{(kn)^k}\fT'_{u+1}(\tau')
\!+\!\ord(n^{-2k})\,.
\label{master-Tutau-1}
\ee
$\ord(1)$ terms disappear leaving $\ord(n^{-k})$ terms from which the following difference-differential equation is deduced:
\be
\frac{\pa\fT'_u(\tau')}{\pa\tau'}=[k(u+1)]^{\ul{k}}\left[\fT'_{u+1}(\tau')-\fT'_u(\tau')\right]\,.
\label{difTutau}\
\ee
Introducing the generating function
\be
\fT'(\mu,\tau')=\sum_{u=0}^\infty \mu^{ku}\fT'_u(\tau')\,,
\label{Tmutau-1}
\ee
one has:
\bea
&\sum_{u=0}^\infty [k(u+1)]^{\ul{k}}\mu^{ku}\fT'_{u+1}(\tau')
=\frac{\pa^k}{\pa \mu^k}\sum_{u=0}^\infty \mu^{k(u+1)}\fT'_{u+1}(\tau')\nonumber\\
&\ \ \ \ \ \ \ \ \ \ \ \ \ \ \ \ \ =\frac{\pa^k}{\pa \mu^k}\left[\sum_{u'=0}^\infty \mu^{ku'}\fT'_{u'}(\tau')-
\fT'_0(\tau')\right]=\frac{\pa^k}{\pa \mu^k}\fT'(\mu,\tau')\nonumber\\
&\sum_{u=0}^\infty [k(u+1)]^{\ul{k}}\mu^{ku}\fT'_u(\tau')
\!=\!\frac{\pa^k}{\pa \mu^k}\sum_{u=0}^\infty \mu^{k(u+1)}\fT'_u(\tau')
\!=\!\frac{\pa^k}{\pa \mu^k}\left[\mu^k\fT'(\mu,\tau')\right].
\label{sums}
\eea
Thus the difference-differential equation \eref{difTutau} can be rewritten as 
a partial-differential equation for the generating function~\eref{Tmutau-1}:
\be
\frac{\pa\fT'(\mu,\tau')}{\pa\tau'}=\frac{\pa^k}{\pa \mu^k}\left[(1-\mu^k)\fT'(\mu,\tau')\right]\,.
\label{pdeTmutau}
\ee

\subsubsection{$\fT'(\mu,\tau')$ for monomers.}

Although it is possible to obtain the probability density
through a direct evaluation of~\eref{Tnut} in the scaling 
limit~\cite{turban15}, here we shall instead deduce it from 
its generating function, $\fT'(\mu,\tau')$, solution of~\eref{pdeTmutau}.
  
For monomers, according to~\eref{tauk} the scaled time
is given by:
\be
\tau'\ev\frac{t}{n}-\ln n\,\qquad -\infty<\tau'<+\infty\,.
\label{tauk1}
\ee
Then according to~\eref{scaltn12}
\be
\ov{\tau'}=\gamma-H_u=-\Psi(u+1)\,,
\label{taumk1}
\ee
where $\Psi(u)$ is 
the digamma function~\cite{lagarias13}.

When $k=1$, the partial differential equation~\eref{pdeTmutau} reads:
\be
\frac{1}{\fT'}\frac{\pa\fT'}{\pa\tau'}=
(1-\mu)\frac{1}{\fT'}\frac{\pa\fT'}{\pa \mu}-1\,.
\label{pde-1}
\ee
The change of function $\fU'=\ln\fT'$ gives:
\be
\frac{\pa\fU'}{\pa\tau'}=(1-\mu)\frac{\pa\fU'}{\pa \mu}-1\,.
\label{pde-2}
\ee
With
\be
\fU'(\mu,\tau')=-\tau'+q(\tau')r(\mu)\,,
\label{Umutau}
\ee
the inhomogeneity is removed and, dividing by $qr$, 
the variables separate
\be
\frac{1}{q}\frac{dq}{d\tau'}=(1-\mu)\frac{1}{r}\frac{dr}{d\mu}=-c_1\,,
\label{pde-3}
\ee
where $c_1$ is a constant. Thus one has
\be\fl
q(\tau')=c_2\e^{-c_1\tau'}\,,\quad
r(\mu)=c_3(1-\mu)^{c_1}\,,\quad\fU'(\mu,\tau')=-\tau'-c'_2(1-\mu)^{c_1}\e^{-c_1\tau'}\,,
\label{qr}
\ee
where $c'_2=-c_2c_3$ is a new constant and
\be
\fT'(\mu,\tau')=\sum_{u=0}^\infty \mu^u\,\fT'_u(\tau')
=\exp\left[-\tau'-c'_2(1-\mu)^{c_1}
\e^{-c_1\tau'}\right]\,.
\label{Tmutau-2}
\ee
$\mu=0$ selects the term $u=0$ in the sum over $u$ so that:
\be
\fT'(0,\tau')=\fT'_0(\tau')=\exp\left[-\tau'-c'_2\e^{-c_1\tau'}\right]\,.
\label{T0tau}
\ee
The integration constants can be determined using
\be\fl 
\int_{-\infty}^\infty\!\fT'(0,\tau')d\tau'\!
=\!\int_{-\infty}^\infty\!\fT'_0(\tau')d\tau'\!=\!1\,,\quad
\int_{-\infty}^\infty\!\!\!\tau'\fT'(0,\tau')d\tau'\!
=\!\int_{-\infty}^\infty\!\!\tau'\fT'_0(\tau')d\tau'\!=\!\ov{\tau'}\!=\!\gamma\,,
\label{norm-mean-1}
\ee
the last relation following from~\eref{taumk1} when $u=0$.
With the change of variable $\theta=\e^{-\tau'}$, one obtains:
\be\fl
\int_{-\infty}^\infty\!\!\fT'(0,\tau')d\tau'\!
\!=\!\int_{0}^\infty\!\!\!\exp\left[-c'_2\theta^{c_1}\right]d\theta\,,\quad
\int_{-\infty}^\infty\!\!\!\tau'\fT'(0,\tau')d\tau'\!
\!=\!-\!\int_{0}^\infty\!\!\ln\theta\exp\left[-c'_2\theta^{c_1}\right]d\theta\,.
\label{norm-mean-2}
\ee
A new change of variable, $\rho=c'_2\theta^{c_1}$, in the first integral
leads to
\be\fl
\int_{-\infty}^\infty\fT'(0,\tau')d\tau'
=\frac{1}{{c'_2}^{1/c_1}c_1}
\int_0^\infty\rho^{1/c_1-1}\e^{-\rho}d\rho
=\frac{\Gamma(1/c_1)}{c_1{c'_2}^{1/c_1}}\,,\quad c'_2
=\left[\frac{\Gamma(1/c_1)}{c_1}\right]^{c_1}\,,
\label{norm}
\ee
where the last relation follows from the normalization of $\fT'_0$ in~\eref{norm-mean-1}.
Inserting this value of $c'_2$ in the second integral of~\eref{norm-mean-2} 
one obtains a standard integral representation of the Euler-Mascheroni 
constant~\cite{lagarias13} when $c_1=1$. 
Thus $c'_2=c_1=1$ and~\eref{Tmutau-2} gives:
\be\fl
\fT'(\mu,\tau')=\exp\left[-\tau'-(1-\mu)\e^{-\tau'}\right]
=\exp\left[-\tau'-\e^{-\tau'}\right]\exp\left[\mu\,\e^{-\tau'}\right]\,.
\label{Tmutau-3}
\ee
The coefficient of $\mu^u$ in the power expansion of the last factor in~\eref{Tmutau-3}
gives the sought probability density, which is a generalized Gumbel distribution~\cite{turban15,chupeau15,pinheiro16}: 
\be
\fT'_u(\tau')=\frac{1}{u!}\exp[-(u+1)\tau'-\e^{-\tau'}]\,,
\quad k=1\,,\quad-\infty<\tau'<+\infty\,.
\label{Tutau-3}
\ee
This result was previously obtained for the covering time by a random walk~\cite{turban15} by taking directly the extreme value scaling limit on $nT_n(s,t)$ with a centered scaling variable, 
$\tau'-\ov{\tau'}$. See also~\cite{baum65} 
where it was noticed that~\eref{Tutau-3} is the $\chi^2$ distribution in the variable $2\e^{-\tau'}$ with $2(u+1)$ degrees of freedom. 

The finite-size data collapse on the standard Gumbel distribution, corresponding to $u=0$, is shown in figure~\ref{fig-7}a.

\subsubsection{$\fT'_u(\tau')$ for dimers.}

\begin{figure}[!t]
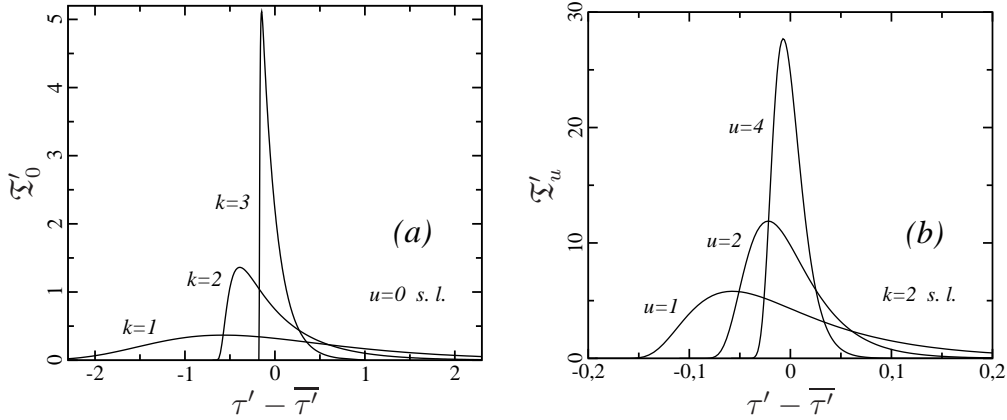

\begin{center}
\psfrag{X0}[Bc][Bc][1][1]{$\fT'_0$}
\psfrag{x}[tc][tc][1][0]{$\tau'-\ov{\tau'}$}
\includegraphics[width=6.5cm,angle=0]{fig-8a.eps}
\hglue 3mm
\psfrag{Xu}[Bc][Bc][1][1]{$\fT'_u$}
\psfrag{x}[tc][tc][1][0]{$\tau'-\ov{\tau'}$}
\includegraphics[width=6.5cm,angle=0]{fig-8b.eps}
\end{center}
\vglue -.5cm
\caption{Evolution of the probability density $\fT'_u(\tau')$ versus 
$\tau'-\ov{\tau'}$ for (a) $u=0$ and increasing values of $k$, 
(b) $k=2$ and increasing values of $u$.  
\label{fig-8}
}
\end{figure}

For dimers, we take the scaling limit directly on the explicit 
expression~\eref{Tnut} with $k=2$. According to~\eref{abcd},
\bea
\fl c&=\prod_{i=j+1}^{n-u-1}2(i-j)[2(i+j)+4u+3]
=\frac{(2n-2u-2j-2)!!(2n+2u+2j+1)!!}{(4u+4j+3)!!}\,,
\nonumber\\
\fl d&=(-1)^j\prod_{i=0}^{j-1}2(j-i)[2(i+j)+4u+3]
=(-1)^j\frac{(2j)!!(4u+4j+1)!!}{(4u+2j+1)!!}\,,
\label{cd-2}
\eea
and~\eref{Tnut} leads to:
\bea\fl
(2n)^{\ul{2}}T_n(n-u,t)=\frac{(2n)!}{(2u)!}\sum_{j=0}^{n-u-1}&
\frac{(-1)^j(4u+4j+3)(4u+2j+1)!!}{(2j)!!(2n-2u-2j-2)!!(2n+2u+2j+1)!!}\nonumber\\
&\ \ \ \ \ \ \ \ \ \ \ \ \ \ \ \ \ \ \ \left\{1-\frac{[2(u\!+\!j+\!1)]^{\ul{2}}}{(2n)^{\ul{2}}}\right\}^{t-1}\,.
\label{2nTnut}
\eea
In the scaling limit with 
\be
\tau'=t/(2n)^{\ul{2}}\,,\qquad 0\leq\tau'<\infty\,,
\label{tauk2}
\ee
such that, according to~\eref{scaltn12},
\be
\ov{\tau'}=\ln 2-H_{2u}+H_u\,,
\label{taumk2}
\ee
one obtains:
\be
\left\{1-\frac{[2(u\!+\!j+\!1)]^{\ul{2}}}{(2n)^{\ul{2}}}\right\}^{t-1}\ev\e^{-(2u+2j+1)(2u+2j+2)\tau'}\,.
\label{exp}
\ee
Using $(2n)!=(2n)!!(2n-1)!!$ the $n$-dependent factors in~\eref{2nTnut} give \footnote{A similar calculation does not seem to be possible for $k=3$ (or more) since $c$ cannot then be expressed as a ratio of triple factorials as in~\eref{cd-2}.}
\be\fl
\frac{(2n)!!/(2n-2u-2j-2)!!}{(2n+2u+2j+1)!!/(2n-1)!!}
=\frac{\overbrace{(2n)(2n-2)\ldots(2n-2u-2j)}^{u+j+1\ \mathrm{factors}}}{\underbrace{(2n+2u+2j+1)\ldots(2n+3)(2n+1)}_{u+j+1\ \mathrm{factors}}}
\ev 1
\label{dblefact}
\ee
in the limit. Thus, for dimers, when $u=n-s=\ord(1)$, the probability density of the covering time is given by:
\bea
\fl\fT'_u(\tau')&\ev(2n)^{\ul{2}}T_n(n-u,t)\nonumber\\
\fl&=\frac{1}{(2u)!}\sum_{j=0}^\infty(-1)^j
\frac{(4u+4j+3)(4u+2j+1)!!}{(2j)!!}
\,\e^{-(2u+2j+1)(2u+2j+2)\tau'}\,.
\label{Tutau-4}
\eea
The finite-size data collapse on this extreme value distribution is shown for $u=0$ in figure~\ref{fig-7}b.

\subsubsection{$\fT'_u(\tau')$ for $k$-mers}
Figure~\ref{fig-8}(a) shows the evolution with $k$ of $\fT'_0$ as a function of $\tau'-\ov{\tau'}$. According to~\eref{Tnut} and~\eref{tauk}, for any value of $k$, the probability density decays exponentially at long time:
\be
\fT'_u(\tau')\sim \exp\left\{-[k(u+1)]^{\ul{k}}\,\tau'\right\}\,,
\qquad \tau'\gg1\,.
\label{asym}
\ee
For increasing values of $u$ the probability density 
evolves from the asymmetric extreme value distribution towards a Gaussian. This is illustrated in figure~\ref{fig-8}(b) for dimers. Similar results can be found for $k=1$ in~\cite{turban15} where the cross-over is studied.

\section{Conclusion}
In this work we have studied the statistics of $k$-mers deposition on the fully-connected lattice. This lattice presents the peculiarity that, for any value of $k$, saturation occurs only at full coverage when the size $N$ is a multiple of $k$. 

In the partial coverage regime, $0<x<1$, where mean-field 
theory is valid, the approach to full coverage is exponential in time for monomers and algebraic for $k>1$. Thus, for dimers, the approach to saturation is algebraic in time on the fully-connected lattice whereas it is exponential on the Cayley tree~\cite{evans84,krapivsky10}.

The probability distribution $T_n(s,t)$ of the time $t$ needed to cover a finite-size system with $s$ $k$-mers, the mean value of the covering time and its variance have been obtained through a generating function approach.

Three different scaling regimes have been studied by taking the scaling limits on the master equations. 

In the low coverage scaling limit ($x\lc0$, 
$s/n^{1/2}\lc y=\ord(1))$,
the probability distribution $D_n(s,v)=T_n(s,s+v)$ converges to a Poisson distribution in the variable $v=t-s$ with mean value and variance $ky^2/2$. 
Thus the random variable $t=s+v$ is self-averaging and a Kronecker delta distribution, $\delta_{v,0}$, is obtained 
when $y\lc0$, i.e., when $s$ grows as $n^\alpha$ 
with $\alpha<1/2$.

In the intermediate coverage scaling limit ($0<x<1$, 
$s=\ord(n)$, $n-s=\ord(n)$), the covering time is self-averaging  with Gaussian fluctuations. The mean value and the variance are growing as $n$. Their scaling functions are both diverging
as $x\to1$, signaling a new scaling behaviour in this limit.
For the mean value
the divergence is logarithmic when $k=1$ and algebraic when $k>1$, whereas it is always algebraic for the variance. 

In the extreme value scaling limit ($x\ev1$, $u=n-s=\ord(1)$)
the mean value of the covering time grows as $n^k$ whereas its variance grows as
$n^{2k}$, thus $t$ is no longer a self-averaging variable. The 
master equation then leads to a difference-differential equation, governing the extreme value distribution of the scaled covering time, for each value of $k$. 

Using a generating function approach this equation has been solved for $k=1$. The fluctuations of the scaled covering time by monomers are governed by a generalized Gumbel distribution. Further work on the solution of the difference-differential equation is needed for $k>1$. For dimers, the scaling limit taken directly on $(2n)^{\ul{2}}T_n$ leads to an alternate infinite series. For any value of $k$ the extreme value distribution displays a $u$- and $k$-dependent exponential decay at long time and crosses 
over slowly to the Gaussian behaviour when $u$ increases.

\ack I thank an anonymous referee for drawing my attention to the coupon collector's problem, particularly to the work on the scaling limits~\cite{baum65}.

\appendix

\section{Inverse of a falling factorial power}
Consider the relation:
\be
A_k(a)=\frac{1}{a^{\ul{k}}}=\frac{1}{\prod_{l=0}^{k-1}(a-l)}
=\frac{1}{(k-1)!}\sum_{l=0}^{k-1}(-1)^{k-l-1}\frac{{k-1\choose l}}{a-l}\,.
\label{aka}
\ee
It is evidently true for $k=1$. Assuming it to be true for $k$, one has:
\be
A_{k+1}(a)=\frac{A_k(a)}{a-k}=\frac{1}{(k-1)!}\sum_{l=0}^{k-1}(-1)^{k-l-1}\frac{{k-1\choose l}}{(a-l)(a-k)}\,.
\label{aka1}
\ee
Using
\be
\frac{1}{(a-l)(a-k)}=\left(\frac{1}{a-k}-\frac{1}{a-l}\right)\frac{1}{k-l}\,,
\label{frac}
\ee
\eref{aka1} can be rewritten as:
\bea
\fl A_{k+1}(a)&=\frac{1}{a-k}\underbrace{\frac{1}{(k-1)!}
\sum_{l=0}^{k-1}(-1)^{k-l-1}\frac{{k-1\choose l}}{k-l}}_{A_k(k)=1/k!}
-\frac{1}{(k-1)!}\sum_{l=0}^{k-1}(-1)^{k-l-1}
\frac{{k-1\choose l}}{(a-l)(k-l)}\nonumber\\
\fl &=\frac{1}{k!(a-k)}+\frac{1}{k!}\sum_{l=0}^{k-1}(-1)^{k-l}
\frac{{k\choose l}}{(a-l)}
=\frac{1}{k!}\sum_{l=0}^{k}(-1)^{k-l}\frac{{k\choose l}}{(a-l)}\,.
\label{aka2}
\eea
Thus \eref{aka} is true for any value of $k$.

\section{From~\eref{tnsk-2} to~\eref{tnsk-3}}

The sum over $l$ in~\eref{tnsk-2} can be rewritten as:
\be\fl
\sum_{l=0}^{k-1}(-1)^{k-l-1}\frac{{k-1\choose l}}{kj-l}
=\underbrace{\sum_{l=0}^{k-1}\frac{1}{kj-l}}_{B_j}+
\underbrace{\sum_{l=0}^{k-1}\frac{(-1)^{k-l-1}{k-1\choose l}-1}{kj-l}}_{C_j}\,.
\label{BjCj
}
\ee
Summing the first term over $j$ gives

\be
\sum_{j=u+1}^nB_j=H_{kn}-H_{ku}\,,
\label{Bj}
\ee
where $H_j=\sum_{i=1}^j1/i$ is a harmonic number.
With 
\be
\frac{1}{kj-l}=\sum_{m=0}^\infty\frac{l^m}{(kj)^{m+1}}
\label{summ-1}\,,
\ee
the last term contributes
\be
\sum_{j=u+1}^n\!\!\!C_j=\sum_{m=0}^\infty\frac{H_n^{(m+1)}-H_u^{(m+1)}}{k^{m+1}}\sum_{l=0}^{k-1}\left[(-1)^{k-l-1}{k-1\choose l}-1\right]l^{m}\,,
\label{Cj-1}
\ee
where $H_j^{(m)}=\sum_{i=1}^j1/i^m$ is a generalized harmonic number.
Since
\be
\sum_{l=0}^{k-1}\left[(-1)^{k-l-1}{k-1\choose l}-1\right]
=\delta_{k,1}-k\,,
\label{deltak1}
\ee
one may separate the contribution from $m=0$ in~\eref{Cj-1}
so that:
\be\fl
\sum_{j=u+1}^n\!\!\!C_j=(\delta_{k,1}-1)(H_n-H_u)+
\sum_{m=2}^\infty\frac{H_n^{(m)}\!-\!H_u^{(m)}}{k^{m}}\sum_{l=0}^{k-1}\left[(-1)^{k-l-1}{k\!-\!1\choose l}\!-\!1\right]l^{m-1}\!\!.
\label{Cj-2}
\ee
Making use of~\eref{Bj} and~\eref{Cj-2} in~\eref{tnsk-2}
finally gives~\eref{tnsk-3}.

\section{From~\eref{var-1} to~\eref{var-2}}
The double sum in~\eref{var-1} may be splitted as:
\bea
\sum_{l,l'=0}^{k-1}(-1)^{l+l'}\frac{{k-1\choose l}{k-1\choose l'}}{(kj-l)(kj-l')} 
&=\underbrace{\sum_{l=0}^{k-1}\frac{1}{(kj-l)^2}}_{D_j}
+\underbrace{\sum_{l=0}^{k-1}
\frac{{k-1\choose l}^2-1}{(kj-l)^2}}_{E_j}\nonumber\\
&+2\underbrace{\sum_{l>l'=0}^{k-1}(-1)^{l+l'}\frac{{k-1\choose l}{k-1\choose l'}}{(kj-l)(kj-l')}}_{F_j}\,.
\label{DEFj}
\eea
Making use of the identity
\be
\frac{1}{(kj-l)^2}=\sum_{m=0}^\infty\frac{(m+1)\,l^m}{(kj)^{m+2}}\,,
\label{summ-2}
\ee
together with~\eref{frac} and~\eref{summ-1}, the sum over $j$ of
the different contributions to~\eref{DEFj} lead to
\bea
\fl\sum_{j=u+1}^nD_j&=H_{kn}^{(2)}-H_{ku}^{(2)}\,,
\nonumber\\
\fl\sum_{j=u+1}^nE_j&=\sum_{m=2}^\infty
\frac{(m-1)\left[H_n^{(m)}-H_u^{(m)}\right]}{k^m}
\sum_{l=0}^{k-1}\left[{k-1\choose l}^2-1\right]l^{m-2}\,,
\nonumber\\
\fl\sum_{j=u+1}^nF_j&=\sum_{m=2}^\infty
\frac{H_n^{(m)}-H_u^{(m)}}{k^m}\sum_{l>l'=0}^{k-1}(-1)^{l+l'}
{k-1\choose l}{k-1\choose l'}\frac{l^{m-1}-{l'}^{m-1}}{l-l'}\,,
\label{sumj}
\eea
where $u=n-s$. Collecting these results in~\eref{var-1} finally 
gives~\eref{var-2}.

\section{Master equation for $\bi{T_n(s,t)}$ in the intermediate coverage scaling limit}
The transition probability~\eref{pns} has the following expansion in powers of $n^{-1}$
\bea
p_n(s-1)&=\prod_{l=0}^{k-1}\left[1-\frac{k(s-1)}{kn-l}\right]
=\prod_{l=0}^{k-1}\left(1-x+\frac{k-lx}{kn}+\ord(n^{-2})\right)
\nonumber\\
&=(1-x)^k+\frac{h(x)}{n}+\ord(n^{-2})\,,
\label{pnsm1}
\eea
where $x=s/n$. Actually the precise form of $h(x)$ is not needed.
The master equation~\eref{master-Tn} can be rewritten 
using~\eref{pnsm1} and $\fT[x(s),\tau(s,t)]$ defined 
in~\eref{Txtau-0} with $\tau$ given by~\eref{tau}. 
A partial differential equation in the variables $\tau$ and $x$ is obtained by expanding $\fT$ to second order in $t$ on the left-hand side and to second order in $s$ on the right-hand side. Higher order derivatives of order $n^{-3/2}$
or more can be neglected. Terms of order 1 give an identity.
At order $n^{-1/2}$ the differential equation
\be
\frac{df}{dx}=-\frac{1}{(1-x)^k}
\label{dftdx}
\ee
is obtained, leading to~\eref{ftx}.
To the next order, $n^{-1}$, the following partial 
differential for the probability density $\fT$ is obtained:
\be
\frac{\pa\fT}{\pa x}\!=\!\frac{1}{2}
\left[\left(\frac{df}{dx}\right)^2\!\!\!-\frac{1}{(1-x)^k}\right]\frac{\pa^2\fT}{\pa\tau^2}\!=\!\frac{1}{2}
\left[\frac{1}{(1-x)^{2k}}\!-\!\frac{1}{(1-x)^k}\right]
\frac{\pa^2\fT}{\pa\tau^2}\,.
\label{pdeT}
\ee

\section{Cross-over from Poisson to Gauss}
When $y$ increases, a cross-over from Poisson to Gauss is observed in finite-size systems (see figure~\ref{fig-4}).
Since $\ov{v}$ and $\ov{\Delta v^2}$ in~\eref{mv} are growing as $y^2$, a continuum approximation in $v$ can be used.
Let us write the Poisson distribution~\eref{Dvy-4} as
\be
\fD_v(y)=e^{F(y,v)}\,,\qquad 
F(y,v)\simeq v(\ln\ov{v}-\ln v)+v-\ov{v}-\ln\sqrt{2\pi v}\,, 
\label{Fvy-1}
\ee
where Stirling's approximation has been used. Expanding $F(y,v)$ to second order in $v$ around its maximum, $v_{max}\sim\ov{v}$, leads to
\be
F(y,v)\simeq F(y,\ov{v})-\frac{(v-\ov{v})^2}{2\ov{v}}\,,
\label{Fvy-2}
\ee
so that
\be
\fD_v(y)\simeq\frac{\exp\left[-\frac{(v-\ov{v})^2}{2\ov{v}}\right]}{\sqrt{2\pi\ov{v}}}\,.
\label{Fvy-3}
\ee
On the Gaussian side, when $x\ll1$, according to~\eref{ftx},~\eref{gtx} and~\eref{tau}, one has:
\be\fl
\ov{t_n}=nf(x)\simeq n\left(x+\frac{kx^2}{2}\right)
=s+\ov{v}\,,\quad 
g(x)\simeq\frac{kx^2}{2}=\frac{\ov{v}}{n}\,,\quad
\tau\simeq\frac{t-s-\ov{v}}{n^{1/2}}
=\frac{v-\ov{v}}{n^{1/2}}\,. 
\label{tngxtau}
\ee
Taking into account~\eref{Txtau-0}, $\fD_v(y)$ has to be compared to $n^{-1/2}\fT(x,\tau)$ so that, making use 
of~\eref{tngxtau} into~\eref{Txtau-2}, a complete agreement with~\eref{Fvy-3} is obtained.

\end{document}


\jl{1}

\title[Random sequential adsorption of k-mers]{Supplement to "Random sequential adsorption of $\bi{k}$-mers on the fully-connected lattice": Statistics of the number of $\bi{k}$-mers on the lattice at a given time}
\author{Lo\"\i c Turban}

\address{Laboratoire de Physique et Chimie Th\'eoriques, Universit\'e de Lorraine--CNRS (UMR7019),
Vand\oe{u}vre l\`es Nancy Cedex, F-54506, France} 

\ead{loic.turban@univ-lorraine.fr}

\begin{abstract} In this supplement to~\cite{turban20}, which is self-contained, we study the probability distribution $S_n(s,t)$ of the number $s$ of adsorbed $k$-mers at time $t$.
We first introduce the discrete master equation governing its evolution
and solve the associated eigenvalue problem. Then we look for the behaviour of $S_n(s,t)$ in two different scaling limits: the short-time limit, $t=\ord(n^{1/2})$, and the intermediate time limit, $t=\ord(n)$.
\end{abstract}
\section{Master equation for $\bi{S_n(s,t)}$}
At each time step $k$ distinct sites are selected at random
among the $N$. Multiple occupancy being forbidden, a new $k$-mer can be adsorbed only when the $k$ selected sites are empty. If $s$ $k$-mers are already covering the lattice, $k(n-s)$ sites among the $kn$ remain free and 
the attempt will be successful with probability
\be
p_n(s)=\frac{{k(n-s)\choose k}}{{kn\choose k}}=\frac{[k(n-s)]^{\ul{k}}}{(kn)^{\ul{k}}}\,,
\label{pns}
\ee
where $j^{\ul{k}}$ is the falling factorial power $j(j-1)\ldots(j-k+1)$~\cite{graham94} and $n$ is the maximum number of adsorbed $k$-mers, corresponding to full coverage.

\begin{figure}[!t]
\begin{center}
\includegraphics[width=8cm,angle=0]{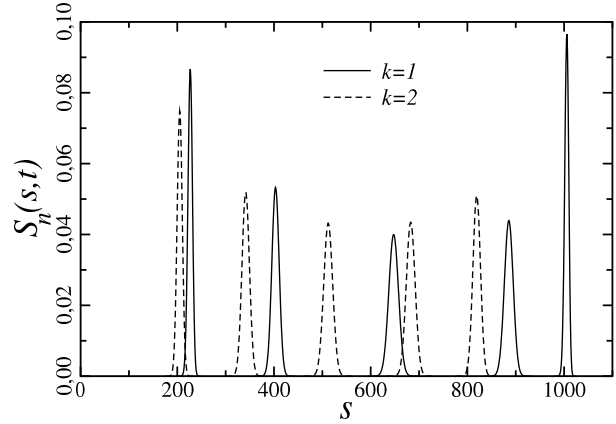}
\end{center}
\vglue -.5cm
\caption{Evolution of $S_n(s,t)$, the probability distribution of the number $s$ of $k$-mers adsorbed at time $t$, obtained by iterating the master equation~\eref{master-Sn}. The lattice size is $2^{10}k$ and the time goes from $2^8$ to $2^{12}$. The width of the distribution has a maximum for intermediate values of $t$. As expected, the deposition rate decreases when 
$k$ or $t$ increases. 
\label{fig-sup-1} 
}
\end{figure}

Let $S_n(s,t)$ be the probability to have $s$ k-mers covering $ks$ sites on the lattice at time $t$ and 
$p_n(s)$ the probability~\eref{pns} to add a k-mer on the lattice with $s$ k-mers. Between $t$ and $t+1$, $s\rightarrow s$ with probability $1-p_n(s)$ and
$s-1\rightarrow s$ with probability $p_n(s-1)$ thus $S_n(s,t)$ evolves according to the following master equation:
\be
S_n(s,t+1)=[1-p_n(s)]S_n(s,t)+p_n(s-1)S_n(s-1,t).
\label{master-Sn}
\ee
A time evolution of the probability distribution is shown in figure~\ref{fig-sup-1}.

The solution of the discrete problem is obtained in the two next sections and its scaling limits are studied in section~\ref{scalSn}.

\section{Solution of the associated eigenvalue problem}

Introducing the column state vector $|S_n(t)\rangle$ with components 
$S_n(s,t)$, $s=0,\ldots,n$, the master equation~\eref{master-Sn} can be rewritten in matrix form as
$|S_n(t+1)\rangle=\mathsf{T}\,|S_n(t)\rangle$ with a transition matrix $\mathsf{T}$ 
given by:
\be
\mathsf{T}=\left(\begin{array}{cccccc}
1\!-\!p_n(0)  &0	     &0	        &0         &0         &0       \\
p_n(0)    &1\!-\!p_n(1)  &0	        &0         &0         &0       \\
	      &  \ddots  &\ddots    &          &          &        \\
0	      &0         &p_n(s\!-\!1)  &1\!-\!p_n(s)  &0         &0       \\
          &          &          &\ddots    &\ddots    &        \\ 
0	      &0	     &0	        & 0	       &p_n(n-1)  &1       \\
\end{array}\right)\,.
\label{tmatrix}
\ee

The eigenvalue equation $\mathsf{T}|\v{\al}\rangle=\lambda_\al|\v{\al}\rangle$ 
leads to the following linear system for the components $\v{\al}_s$ of 
the column eigenvector $|\v{\al}\rangle$
\be
p_n(s-1)\v{\al}_{s-1}+[1-p_n(s)]\v{\al}_{s}=\lambda_\al\v{\al}_s\,,\quad s=0,\cdots,n\,,
\label{syst}
\ee
with $p_n(-1)=p_n(n)=0$. Introducing the eigenvalues 
\be
\lambda_\al=1-p_n(\al)\,,\qquad \al=0,\cdots,n\,,
\label{eval}
\ee
into~\eref{syst}, one obtains a recursion relation 
for the components of the eigenvectors:
\be
\v{\al}_{s}=\frac{p_n(s-1)}{p_n(s)-p_n(\al)}\,\v{\al}_{s-1}\,.
\label{rec}
\ee
Iterating this relation leads to:
\be
\v{\al}_s=\left\{
\begin{array}{ccc}
 \v{\al}_\al\,\prod_{\bet=\al+1}^{s}\frac{p_n(\bet-1)}{p_n(\bet)-p_n(\al)}\,,\qquad &s>\al\\
 \ms
 0\,,\qquad &s<\al
\end{array}
\right.,
\label{sol1}
\ee
The unknown $\v{\al}_\al$ will be fixed by the initial conditions.

\section{General expression of $\bi{S_n(s,t)}$}
The probability distribution of $s$, the number of $k$-mers on the lattice at time
$t$, is obtained by a repeated application of the transition matrix $\mathsf{T}$
on the initial state vector:
\be 
|S_n(t)\rangle=\mathsf{T}^t |S_n(0)\rangle\,.
\label{Snt1}
\ee
Writing the initial state vector as
\be
|S_n(0)\rangle=\sum_{\alpha=0}^n|\v{\al}\rangle\,,
\label{Sn0}
\ee
one obtains
\be
|S_n(t)\rangle=\sum_{\alpha=0}^n \lambda_\al^t|\v{\al}\rangle\,,
\label{Snt2}
\ee
which in components, according to~\eref{sol1}, translates into:
\be
S_n(s,t)=\sum_{\alpha=0}^s \v{\al}_s\lambda_\al^t\,.
\label{Snst1}
\ee
Assuming an empty lattice at $t=0$, the initial condition 
reads:
\be
S_n(s,0)=\sum_{\alpha=0}^s \v{\al}_s=\delta_{s,0}\,.
\label{Sns0}
\ee
The unknown $\v{\al}_\al$, which follow from the last relation (see appendix A),
are given by
\be
\v{\al}_\al=\prod_{\bet=0}^{\al-1}\frac{p_n(\bet)}{p_n(\bet)-p_n(\al)}\,,
\label{valal}
\ee
so that \eref{sol1} can be rewritten as:
\be
\v{\al}_s=\left\{
\begin{array}{ccc}
 \frac{\prod_{\bet=0}^{s-1}p_n(\bet)}{\prod_{\bet=0\atop \bet\neq\al}^{s}[p_n(\bet)-p_n(\al)]}\,,\qquad &s\geq\al\\
 \ms
 0\,,\qquad &s<\al
\end{array}
\right.,
\label{sol2}
\ee
Finally, making use of~\eref{eval} and \eref{sol2} in~\eref{Snst1},
one obtains the probability distribution:
\be
S_n(s,t)=\prod_{\bet=0}^{s-1}p_n(\bet)\sum_{\alpha=0}^s 
\frac{[1-p_n(\al)]^t}{\prod_{\bet=0\atop \bet\neq\al}^s[p_n(\bet)-p_n(\al)]}\,.
\label{Snst2}
\ee
For $k=1$ and $p_n(s)=1-s/n$ this probability distribution can be written as~\cite{turban14}
\be
S_n(s,t)=\frac{n^{\ul{s}}}{n^t}{t\brace s}
\label{Snstir-1}
\ee
where
\be
{t\brace s}=\frac{1}{s!}
\sum_{j=0}^{s}(-1)^{s-j}{s\choose j}j^{t}
\label{Snstir-2}
\ee
is a Stirling number of the second kind~\cite{stirling49}.

Although analytical results for the mean 
value $\ov{s_n}$ and the variance 
$\ov{\Delta s_n^2}$ can be deduced from~\eref{Snstir-1} for monomers~\cite{turban14}, it seems difficult to do the same
for higher values of $k$. It turns out that these calculations 
are much easier in the different scaling limits, 
as shown in the next section.

\section{Scaling limits for $\bi{S_n(s,t)}$\label{scalSn}}
\subsection{Short-time scaling limit}
\begin{figure}[!t]
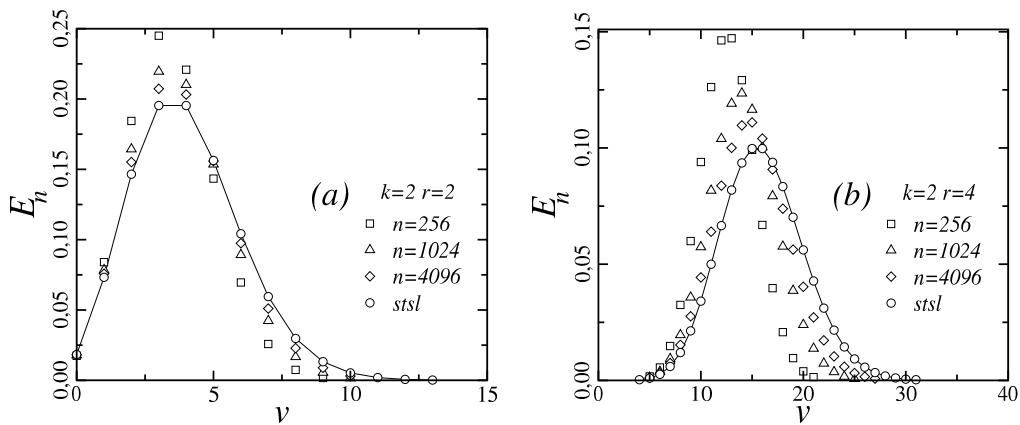

\begin{center}
\includegraphics[width=6.5cm,angle=0]{fig-sup-2a.eps}
\hglue .3cm
\includegraphics[width=6.5cm,angle=0]{fig-sup-2b.eps}
\end{center}
\vglue -.5cm
\caption{Evolution of the finite-size data for $E_n(v,t)$ vs $v=t-s$ towards the Poisson distribution~$\fE_v(r)$ in~\eref{fEv-4} for dimers when (a) $r=2$ and (b) $r=4$. There is a crossover to a Gaussian behaviour when $n$ decreases or $r$ increases.
\label{fig-sup-2} 
}
\end{figure}

We first consider a {\it short-time scaling limit} (stsl) suggested by the low-coverage scaling limit first considered in~\cite{baum65}. In this limit $s,t,n\to\infty$ with $t=\ord(n^{1/2})$ so that:
\be
r=\frac{t}{n^{1/2}}\stsl\ord(1)\,,\qquad 
w=\frac{t}{n}=\frac{r}{n^{1/2}}\stsl0\,.
\label{r}
\ee
At short time most lattice sites are empty and almost 
every deposition attempt is successful, so the relevant 
variable is rather $v=t-s$ than $s$ itself. For $E_n(v,t)$ given by $S_n(s,t)$ in~\eref{master-Sn}
one obtains the following evolution: 
\be
E_n(v,t+1)=[1-p_n(s+1)]E_n(v-1,t)+p_n(s)E_n(v,t)\,.
\label{master-En}
\ee
Let us look for the form taken by this master equation in the short-time scaling limit where we define:
\be
\fE_v(r)=\fE_v[r(t)]\stsl E_n(v,t)\,,\qquad 
v\in\mathbb{N}_0\,.
\label{fEv-0}
\ee
On the left-hand side we have:
\be 
\fE_v[r(t+1)]=\fE_v(r+n^{-1/2})=\fE_v(r)+
\frac{1}{n^{1/2}}\frac{\pa\fE_v}{\pa r}+\ord(n^{-1})\,,
\label{fEv-1}
\ee 
On the right the transition probability~\eref{pns} takes the following form:
\be\fl 
p_n(s)=p_n(rn^{1/2}-v)=\prod_{j=0}^{k-1}\frac{k(n-rn^{1/2}+v)-j}{kn-j}=1-\frac{kr}{n^{1/2}}+\ord(n^{-1})=p_n(s+1)\,.
\label{pnr}
\ee
Thus the master equation~\eref{master-En} leads to:
\be 
\fE_v(r)+\frac{1}{n^{1/2}}\frac{\pa\fE_v}{\pa r} 
=\frac{kr}{n^{1/2}}\fE_{v-1}(r)+
\left(1-\frac{kr}{n^{1/2}}\right)\fE_v(r)+\ord(n^{-1})\,.
\label{fEv-2}
\ee
To leading order, $n^{-1/2}$, one obtains the following difference-differential equation for $\fE_v(r)$:
\be 
\frac{\pa\fE_v}{\pa r} 
=kr\left[\fE_{v-1}(r)-\fE_v(r)\right]\,.
\label{fEv-3}
\ee
The normalized solution is the Poisson distribution
\be 
\fE_v(r)=\frac{(kr^2/2)^v}{v!}\exp\left(\frac{-kr^2}{2}\right)
\label{fEv-4}
\ee
so that:
\be 
\ov{v}=\frac{kr^2}{2}\stsl t-\ov{s_n}\,,\qquad
\ov{\Delta v^2}=\frac{kr^2}{2}\stsl\ov{\Delta s_n^2}\,.
\label{moy}
\ee
When $t=\ord(n^\alpha)$ with $\alpha<1/2$, according to~\eref{r}, $r$ vanishes in the short-time scaling limit. Then the Poisson distribution reduces to a Kronecker delta distribution, $\fE_v(0)=\delta_{v,0}$. 

The evolution of the finite-size data towards the Poisson distribution  
is shown in figure~\ref{fig-sup-2} for dimers.

\subsection{Intermediate time scaling limit}

\begin{figure}[!t]
\begin{center}
\includegraphics[width=8cm,angle=0]{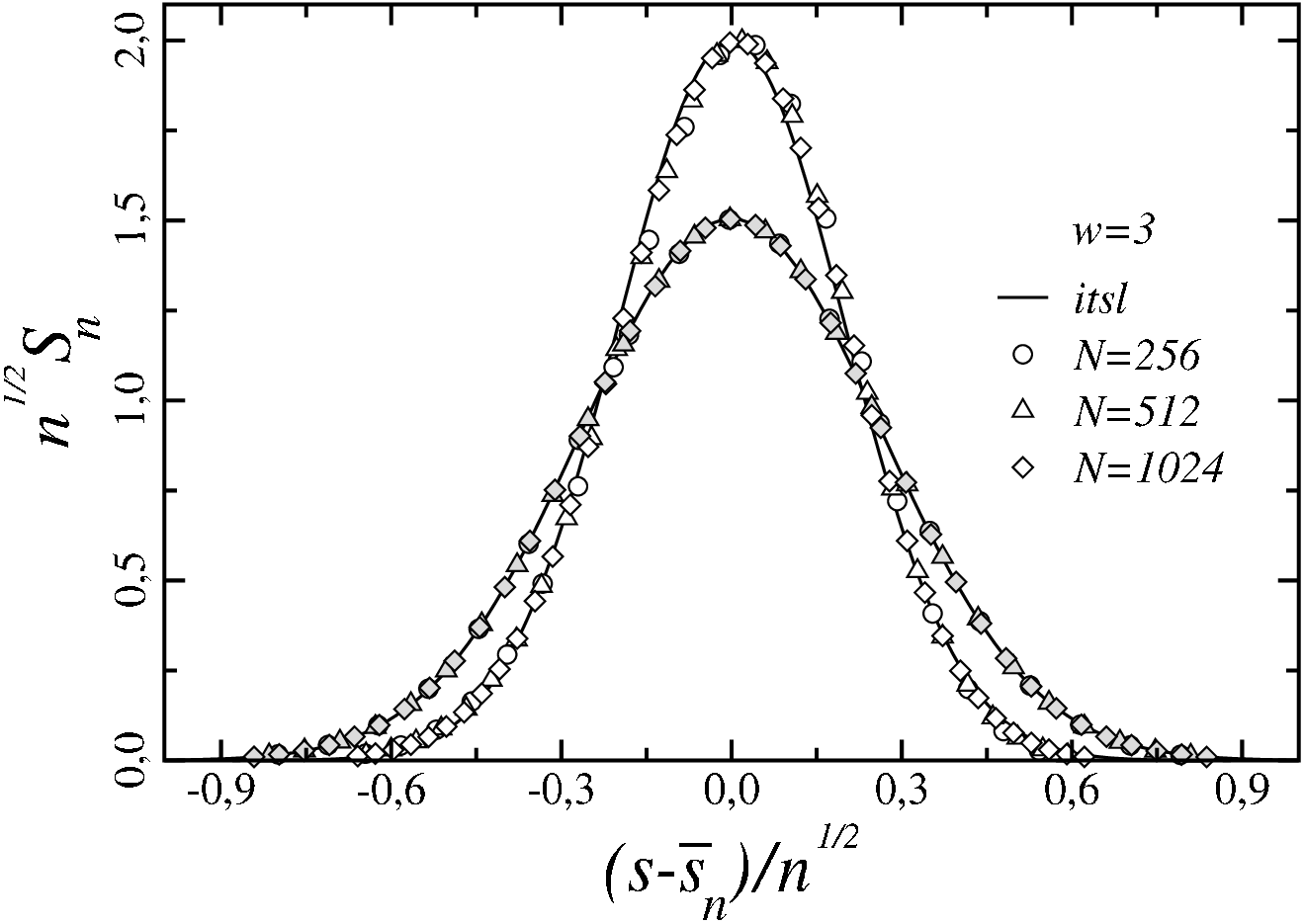}
\end{center}
\vglue -.5cm
\caption{Finite-size data collapse on the Gaussian scaling function 
$\fS\itsl n^{1/2}S_n$ in~\eref{Ssigmaw-2} versus 
$\sigma=(s-\ov{s_n})/n^{1/2}$
for $k=1$ (white symbols), $k=2$ (grey symbols) when $w=t/n=3$.  
\label{fig-sup-3}
}
\end{figure}

Let us now consider the master 
equation~\eref{master-Sn} in the 
{\it intermediate time scaling limit} (itsl), when $t$ and $n$ are sent to infinity with a 
fixed value of their ratio, $w=t/n$. To do so, we have to specify 
the expression of the transition probability, $p_n(s)$,
which, for $k$-mers, is given by~\eref{pns}. 
Mean-field theory~\cite{turban20} suggests that the mean number $\ov{s_n}$ of $k$-mers at time $t$ scales as $n$ and is a function of $w$. Assuming a similar behaviour for the variance, we introduce the 
unknown scaling functions 
$f_s(w)$ and $g_s(w)$ such that:
\be
\frac{\ov{s_n}}{n}\itsl f_s(w)\,,\qquad 
\frac{\ov{\Delta s_n^2}}{n}\itsl g_s(w)\,,
\qquad w\itsl\frac{t}{n}\,.
\label{fsgs}
\ee
Introducing the scaled and centered variable 
\be 
\sigma(s,t)\itsl\frac{s-\ov{s_n}}{n^{1/2}}=\frac{s}{n^{1/2}}-n^{1/2}f_s
\left(\frac{t}{n}\right)\,,
\label{sigma}
\ee
the probability density, in the continuum limit, is given by
(see figure~\ref{fig-sup-3}):
\be
\fS(\sigma,w)=\fS[\sigma(s,t),w(t)]\itsl n^{1/2}S_n(s,t)\,.
\label{Ssigmaw-1} 
\ee
\begin{figure}[!tb]
\begin{center}
\includegraphics[width=6.5cm,angle=0]{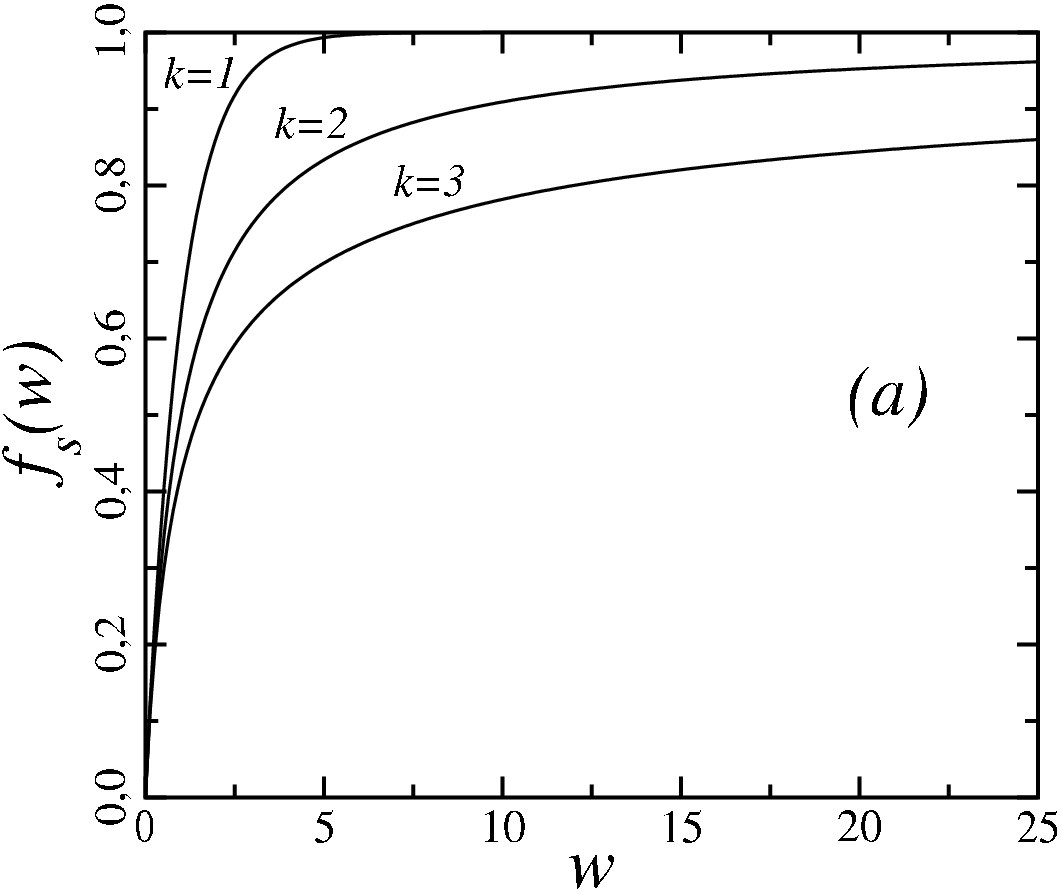}
\hglue .3cm
\includegraphics[width=6.5cm,angle=0]{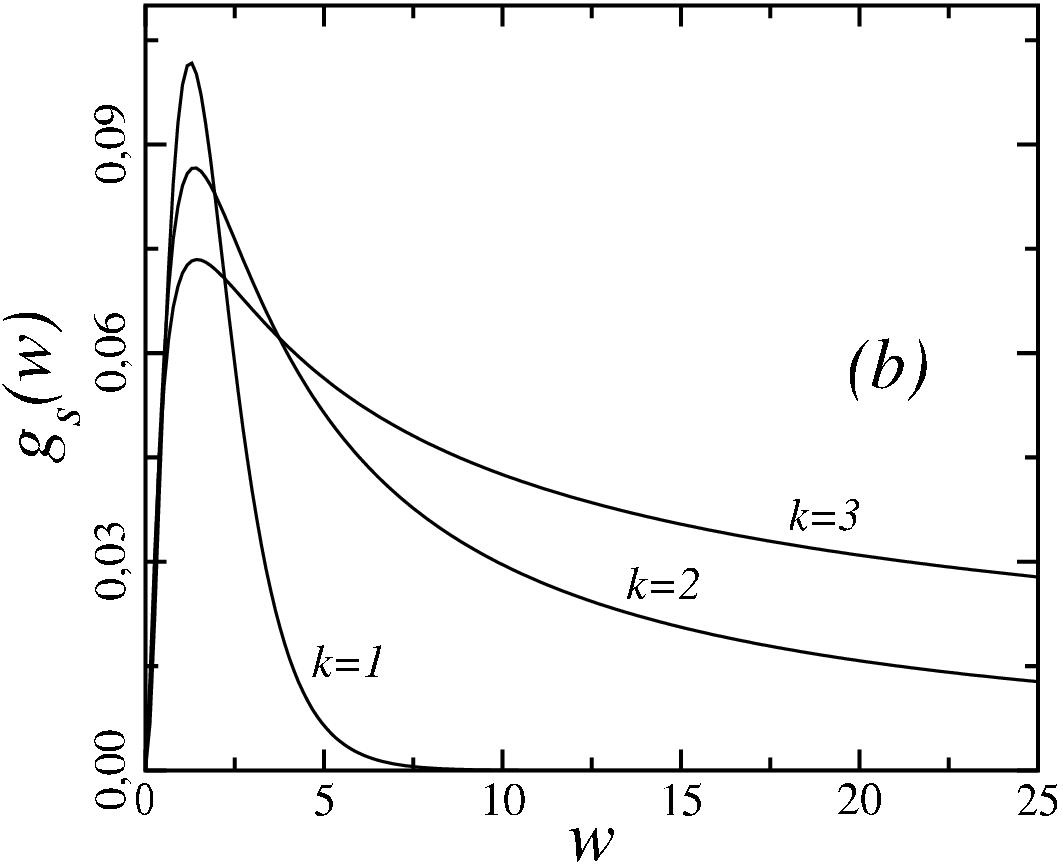}
\end{center}
\vglue -.5cm
\caption{Scaling functions (a) $f_s(w)\itsl\ov{s_n}/n$ and (b)
$g_s(w)\itsl\ov{\Delta s_n^2}/n$ versus $w=t/n$ for $k=1,2,3$.
The variance vanishes when $w\to0$ and when $w\to\infty$. 
In these limits the probability density $\fS$ is a Dirac 
$\delta$ function. In both cases the long time asymptotic value is approached exponentially when $k=1$ and with a $k$-dependent power when $k>1$. Note the change of vertical scale.
\label{fig-sup-4} 
}
\end{figure}

Expanding the master equation~\eref{master-Sn} (see appendix B), terms of order
$n^{-1/2}$ give a differential equation~\eref{dfsdw} for $f_s(w)$. The solution with $f_s(0)=0$ reads
\be
f_s(w)\itsl\frac{\ov{s_n}}{n}=\left\{
\begin{array}{ccc}
 1-[1+(k-1)w]^{-1/(k-1)}\,,\qquad &k>1\\
 \ms
1-e^{-w} \,,\qquad &k=1
\end{array}
\right.\,.
\label{fsw}
\ee
in agreement with the mean-field result in~\cite{turban20}.
The scaling function is shown in figure~\ref{fig-sup-4}(a). When $w\to0$, $f_s(w)$ vanishes as $w-kw^2/2$. The linear term is independent of $k$, as expected for the deposition of 
a $k$-mer at each time step. The quadratic correction leads to~$\ov{s_n}=t-kr^2/2$ as given by~\eref{moy} in the short-time scaling limit.
The distance to saturation $1-f_s(w)$ vanishes exponentially when $k=1$ and as a $k$-dependent power, $w^{-1/(k-1)}$ when $k>1$. 

To order $n^{-1}$ a partial differential equation for $\fS(\sigma,w)$ 
is obtained in~\eref{pdeS}. Rewriting the probability density as 
\be
\fS(\sigma,w)=\fP[\sigma,g_s(w)]\,,
\label{fsfp}
\ee
where $g_s(w)$ is the unknown scaling function of the variance in~\eref{fsgs},
leads to
\bea
\frac{\pa\fP}{\pa g_s}\frac{dg_s}{dw}&=\frac{1}{2}
\underbrace{\left[(1-f_s)^k-(1-f_s)^{2k}-2k(1-f_s)^{k-1}g_s\right]}_{=dg_s/dw}
\frac{\pa^2\fP}{\pa\sigma^2}\nonumber\\
&\ \ \ \ \ \ \ \ \ \ \ \ \ \ \ \ \ \ \ \ 
+k(1-f_s)^{k-1}\underbrace{\left[\fP+\sigma\frac{\pa\fP}{\pa\sigma}+g_s
\frac{\pa^2\fP}{\pa\sigma^2}\right]}_{=0}\,.
\label{pdeP}
\eea
where the last term subtracted in the first bracket is added in the last one.
Then the last bracket vanishes for a Gaussian with variance $g_s$
\be
\fS(\sigma,w)=\fP(\sigma,g_s)=\frac{\exp\left[-\frac{\sigma^2}{2g_s(w)}
\right]}{\sqrt{2\pi g_s(w)}}
\label{Ssigmaw-2}
\ee
which is the solution of the diffusion equation, obtained self-consistently
when $g_s$ satisfies the differential equation
\be
\frac{dg_s}{dw}=(1-f_s)^k-(1-f_s)^{2k}-2k(1-f_s)^{k-1}g_s\,.
\label{dgsdw}
\ee
As shown in appendix C, the solution vanishing when $w=0$ 
reads:
\be\fl
g_s(w)\itsl\frac{\ov{\Delta s_n^2}}{n}=\left\{
\begin{array}{ccc}
 \frac{1}{2k-1}\left\{\frac{1}{[1+(k-1)w]^{1/(k-1)}}
 -\frac{1+(2k-1)w}{[1+(k-1)w]^{2k/(k-1)}}\right\}
 \,,\qquad  &k>1\\
 \ms
e^{-w}-(1+w)e^{-2w}\,, &k=1
\end{array}
\right..
\label{gsw}
\ee
This scaling function is shown in figure~\ref{fig-sup-4}(b). When $w\to0$ the variance $g_s(w)$ vanishes as $kw^2/2$
so that $\ov{\Delta s_n^2}\simeq kr^2/2$, in agreement with~\eref{moy} in the short-time scaling limit.
When $w\gg1$ it behaves like $1-f_s(w)$, vanishing exponentially when $k=1$ and as $w^{-1/(k-1)}$ when $k>1$. Thus the Gaussian 
probability density, $\fS(\sigma,w)$ in~\eref{Ssigmaw-2}, becomes a Dirac $\delta$ function in both limits.

\appendix

\section{Evaluation of $\bi{\v{\al}_\al}$}
For $s=0$ the initial condition in~\eref{Sns0} gives 
$\v{0}_0=1$. When $s=1$, $\v{0}_1+\v{1}_1=0$ and,
using~\eref{sol1}, one obtains:
\be
\v{1}_1=\frac{p_n(0)}{p_n(0)-p_n(1)}\,.
\label{v11}
\ee
For $s=2$, $\v{0}_2+\v{1}_2+\v{2}_2=0$, so that  
\be
\v{0}_2=-\frac{p_n(0)}{p_n(0)-p_n(1)}
\times\frac{p_n(1)}{p_n(0)-p_n(2)}\times1
\label{v20}
\ee
and
\be
\v{1}_2=-\frac{p_n(1)}{p_n(1)-p_n(2)}
\times\frac{p_n(0)}{p_n(0)-p_n(1)}
\label{v21}
\ee
lead to:
\be
\v{2}_2=\frac{p_n(0)p_n(1)}{[p_n(0)-p_n(2)][p_n(1)-p_n(2)]}\,.
\label{v22}
\ee
These results suggest the general expression~\eref{valal}
leading to~\eref{sol2}. 

Let us now prove it by induction. Assuming that~\eref{valal} (and, consequently~\eref{sol2}) is true 
for $\al\leq s-1$, the initial condition in~\eref{Sns0} gives:
\bea
\v{s}_s&=-\sum_{\al=0}^{s-1}\v{\al}_s=-\sum_{\al=0}^{s-1}
\frac{\prod_{\bet=0}^{s-1}p_n(\bet)}{\prod_{\bet=0\atop \bet\neq\al}^{s}[p_n(\bet)-p_n(\al)]}
\nonumber\\
&=-\underbrace{\sum_{\al=0}^{s}
\frac{\prod_{\bet=0}^{s-1}p_n(\bet)}{\prod_{\bet=0\atop \bet\neq\al}^{s}[p_n(\bet)-p_n(\al)]}}_{F_s}
+\prod_{\bet=0}^{s-1}\frac{p_n(\bet)}{p_n(\bet)-p_n(s)}\,.
\label{vss}
\eea
In order to evaluate $F_s$, let us introduce the following auxiliary function of the complex variable $z$: 
\be
F(z)=\frac{\prod_{\bet=0}^{s-1}p_n(\bet)}{\prod_{\bet=0}^{s}[z-p_n(\bet)]}\,.
\label{fz}
\ee
It has $s+1$ simple poles $z_\al=p_n(\al)$, $\al=0,\cdots,s$ and vanishes as 
$|z|^{-s-1}$ when $|z|\to\infty$. Thus the integral of $F(z)$ on a circle with radius $R$ centered at the origin vanishes when $R\to\infty$. As a consequence, the sum of the residues at the $s+1$ poles, given by $F_s$ up to a sign, vanishes too and the last term in~\eref{vss} gives $\v{s}_s$ in agreement with the expected form~\eref{valal}.

\section{Master equation for $\bi{S_n(s,t)}$ in the scaling limit}
According to~\eref{sigma} one has
\be
s=n^{1/2}\sigma+nf_s(w)
\label{s}
\ee
which can be used in~\eref{pns} to give
\be\fl
p_n(s)=(1\!-\!f_s)^k-\frac{k}{n^{1/2}}\sigma(1\!-\!f_s)^{k\!-\!1}
+\frac{k(k\!-\!1)}{2n}\left[\sigma^2(1\!-\!f_s)^{k-2}\!-\!f_s(1\!-\!f_s)^{k\!-\!1}\right]
+\ord(n^{-3/2})\,,
\label{pnsigma-1}
\ee
and
\be
p_n(s-1)=p_n(s)+\frac{k}{n}(1-f_s)^{k-1}\,.
\label{pnsigma-2}
\ee
The master equation in~\eref{master-Sn} can be rewritten using these relations and $\fS[\sigma(s,t),w(t)]$ in~\eref{Ssigmaw-1}. A partial differential equation in the variables $\sigma$ and $w$ is obtained by expanding $\fS$ to second order in $t$ on the left-hand side and to second order in $s$ on the right-hand side. Higher order derivatives of order $n^{-3/2}$
or more can be neglected. Terms of order 1 give an identity. At order
$n^{-1/2}$ the differential equation
\be
\frac{df_s}{dw}=(1-f_s)^k
\label{dfsdw}
\ee
is obtained, leading to~\eref{fsw}. The next order, $n^{-1}$, gives the following partial differential equation for the probability density:
\be
\frac{\pa\fS}{\pa w}=\frac{1}{2}
\left[(1-f_s)^k-(1-f_s)^{2k}\right]\frac{\pa^2\fS}{\pa\sigma^2} 
+k(1-f_s)^{k-1}\left[\fS+\sigma\frac{\pa\fS}{\pa\sigma}\right]\,.
\label{pdeS}
\ee

\section{Solution of the differential equation for $\bi{g_s(w)}$}
The solution of~\eref{dgsdw} is easily obtained using the variation of constants method. Using~\eref{fsw} one obtains
\be
1-f_s(w)=\frac{1}{[1+(k-1)w]^{1/(k-1)}}
\label{fsw-1}
\ee
the homogeneous equation takes the following form
\be
\frac{dg_s}{dw}+\frac{2kg_s}{1+(k-1)w}=0\,,
\label{difgs}
\ee
so that:
\be
g_s=\frac{c}{[1+(k-1)w]^{2k/(k-1)}}\,.
\label{gs}
\ee
Introducing this expression in~\eref{dgsdw} leads to a differential
equation for $c(w)$
\be
\frac{dc}{dw}=[1+(k-1)w]^{k/(k-1)}-1\,,
\label{dcdw}
\ee
giving:
\be
c(w)=\frac{1}{2k-1}[1+(k-1)w]^{(2k-1)/(k-1)}-w-d\,.
\label{cw-1}
\ee
The initial condition $g_s(0)=0$ implies $c(0)=0$ and $d=-1/(2k-1)$
so that:
\be
c(w)=\frac{1}{2k-1}\{[1+(k-1)w]^{(2k-1)/(k-1)}-1\}-w\,.
\label{cw-2}
\ee
Combining this result with~\eref{gs} gives the first 
expression in~\eref{gsw}. The second is obtained in the limit $k\to 1$.